\documentclass[pdflatex,sn-nature]{sn-jnl}
\usepackage{graphicx} % Required for inserting images
\usepackage{multirow}
\usepackage{booktabs}
\usepackage{amsmath,amssymb,amsfonts}%
\usepackage{amsthm}%
\usepackage{mathrsfs}%
\usepackage[title]{appendix}%
\usepackage{xcolor}%
\usepackage{textcomp}%
\usepackage{manyfoot}%
\usepackage{algorithm}%
\usepackage{algorithmicx}%
\usepackage{algpseudocode}%
\usepackage{listings}%

\theoremstyle{thmstyleone}%
\theoremstyle{thmstyletwo}%
\theoremstyle{thmstylethree}%

\begin{document}

\title[Article Title]{Large Model Driven Solar Activity AI Forecaster: A Scalable Dual Data-Model Framework}

\author[1]{\fnm{Jingjing} \sur{Wang}} 
\equalcont{These authors contributed equally to this work.}

\author[2]{\fnm{Pengyu} \sur{Liang}}
\equalcont{These authors contributed equally to this work.}

\author[1]{\fnm{Tingyu} \sur{Wang}}
\equalcont{These authors contributed equally to this work.}

\author[1]{\fnm{Ming} \sur{Li}}

\author[1]{\fnm{Yanmei} \sur{Cui}}

\author[1,3]{\fnm{Siwei} \sur{Liu}}

\author[4]{\fnm{Xin} \sur{Huang}}

\author[2]{\fnm{Xiang} \sur{Li}}

\author[2]{\fnm{Minghui} \sur{Zhang}}

\author[1,3]{\fnm{Yunshi} \sur{Zeng}}

\author[1,3]{\fnm{Zhu} \sur{Cao}}

\author[2]{\fnm{Jiekang} \sur{Feng}}

\author*[2]{\fnm{Qinghua} \sur{Hu}}\email{huqinghua@tju.edu.cn}

\author*[1,3]{\fnm{Bingxian} \sur{Luo}}\email{luobx@nssc.ac.cn}

\author*[2]{\fnm{Bing} \sur{Cao}}\email{caobing@tju.edu.cn}

\affil[1]{\orgname{State Key Laboratory of Solar Activity and Space Weather, National Space Science Center, Chinese Academy of Sciences}, \orgaddress{\street{NO.1 Nanertiao, Zhongguancun, Haidian district}, \city{Beijing}, \postcode{100190}, \country{China}}}

\affil[2]{\orgdiv{School of Artificial Intelligence}, \orgname{Tianjin University}, \orgaddress{\street{No. 135 Yaguan Road}, \city{Tianjin}, \postcode{300350}, \country{China}}}

\affil[3]{\orgname{University of Chinese Academy of Sciences}, \orgaddress{\street{No.1 Yanqihu East Rd, Huairou District}, \city{Beijing}, \postcode{100049}, \country{China}}}

\affil[4]{\orgdiv{Faculty of Electrical Engineering and Computer Science}, \orgname{Ningbo University}, \orgaddress{\street{No.818 Fenghua Road}, \city{Ningbo}, \postcode{610101}, \country{China}}}

\abstract{Solar activity drives space weather, affecting Earth's magnetosphere and technological infrastructure, which makes accurate solar flare forecasting critical. Current space weather models under-utilize multi-modal solar data, lack iterative enhancement via expert knowledge, and rely heavily on human forecasters under the Observation-Orientation-Decision-Action (OODA) paradigm. Here we present the "Solar Activity AI Forecaster", a scalable dual data-model driven framework built on foundational models, integrating expert knowledge to autonomously replicate human forecasting tasks with quantifiable outputs. It is implemented in the OODA paradigm and comprises three modules: a Situational Perception Module that generates daily solar situation awareness maps by integrating multi-modal observations; In-Depth Analysis Tools that characterize key solar features (active regions, coronal holes, filaments); and a Flare Prediction Module that forecasts strong flares ($\geq$M-class) for the full solar disk and active regions. Executed within a few minutes, the model outperforms or matches human forecasters in generalization across multi-source data, forecast accuracy, and operational efficiency. This work establishes a new paradigm for AI-based space weather forecasting, demonstrating AI’s potential to enhance forecast accuracy and efficiency, and paving the way for autonomous operational forecasting systems.}


\maketitle

\section{Introduction}\label{intro}

\textbf{Solar eruption}—encompassing flares, coronal mass ejections (CMEs), and solar proton events (SPEs)—constitute the main drivers of space weather \cite{gonzalez1999gsorigin,leka2018origin}, with cascading effects on Earth’s magnetosphere, technological infrastructure \cite{eastwood2017economicimpact}, and human activities \cite{schwenn2006sw,tsurutani2009affects,baker2013extreme,marov2021solar,temmer2021sw}. These phenomena are governed by magnetic reconnection processes in twisted flux ropes \cite{amari2018magrope,cairns2018lowsolarreconnect} and propagate through highly structured solar wind \cite{wang1996chmagnature,bale2019chslowsw}, inducing geomagnetic storms that disrupt radio communications and jeopardize satellite operations, with extreme events posing significant risks to power grids and economic stability \cite{oughton2017economicimpact}. The magnetic nature of coronal structures further modulates eruption dynamics, necessitating accurate forecasting for risk mitigation.

In solar physics research, accurate identification of solar features (active region, coronal hole, and filament) is fundamental to gaining an in-depth understanding of solar activity patterns, and advances in solar monitoring technology and machine learning have provided strong support for this \cite{huang2024review,2024SoPh..299..121G,2025ApJS..278...53H,GRYCUK2025102604,2015JSWSC...5A..23R,2020SoPh..295..110B,2021A&A...652A..13J,2019arXiv191202743A,2014A,2021SoPh..296..176L,2024A&A...686A.213D,2025ApJ...980..176Z,https://doi.org/10.1029/2023SW003516}. For example, Grycuk \cite{GRYCUK2025102604} proposed a fast detection method based on the Layer-Sector Solar Hash (LSSH), which utilizes SDO/AIA EUV images to improve the retrieval speed by two orders of magnitude while maintaining robustness to solar rotation. Jarolim \cite{2021A&A...652A..13J} developed a convolutional neural network, CHRONNOS, which combines EUV and magnetic field data to automate coronal hole detection with 98$\%$ accuracy. Diercke et al.  \cite{2024A&A...686A.213D} used ChroTel, GONG, and KSO H$\alpha$ full-disk images and a hybrid deep learning framework combining YOLOv5 detection and U-Net segmentation to achieve automatic detection, pixel-level segmentation, and extraction of physical parameters (such as area and inclination angle) of filaments. Since the detected features are used for later forecasting, identifying solar features is really important. Most of the current feature detection work relies on a single data source, such as using magnetograms to detect active regions, EUV data to detect coronal holes, and H$\alpha$ images to detect filaments. 

Building on the effective identification of solar activity features, further research on flare prediction has become an important direction in the field of solar physics, with photospheric magnetic parameters playing a key role. Photospheric magnetograms and magnetic parameters derived from space-weather HMI Active Region Patches (SHARPs) serve as pivotal predictors, particularly those characterizing non-potential magnetic fields \cite{leka2007parameters,schrijver2007rvalue,bobra2014sharp,wang2019pilm}. Magnetograms provide high-resolution spatial distribution information, clearly revealing the fine structures of solar active regions, while magnetic field characteristics offer interpretable physical quantities such as magnetic field strength and gradients, thereby enhancing the physical interpretability of models. Both types of data are highly effective for flare prediction. Although numerous artificial intelligence-based flare forecasting studies have been conducted \cite{2023NatSR..1313665A,2024SoPh..299...33G,2020SpWea..1802440J,
2022ApJS..258...12L,2018ApJ...858..113N,2018ApJ...869...91P,2024FrASS..1098609P,2022ApJ...941....1S,2021RAA....21..160A,2023FrASS...939805G,huang2018deep,2019ApJ...877..121L,2022A&A...662A.105G}, most utilize only one modality—either images or physical parameters—to construct flare prediction models, significantly limiting further improvements in forecasting accuracy. In recent years, some studies have employed both magnetograms and physical parameters to build forecasting models \cite{li2022knowledge,2021A&C....3500468R,2022ApJ...931..163S,2021ApJS..257...50T}. However, due to inherent limitations in model architecture, these approaches have failed to effectively capture and integrate deep correlations among different modal features, resulting in limited performance in complex tasks \cite{2025RAA....25c5025H}. Additionally, affected by projection effects, many studies typically restrict their analysis to the central longitude range of the solar disk, failing to meet current operational flare-propagation requirements. Therefore, to effectively integrate multi-modal features and address operational forecasting needs, employing large models to achieve deeper cross-modal feature interaction represents a viable approach. Meanwhile, machine learning has enabled novel applications in solar physics, such as AI-reconstructed farside magnetograms \cite{kim2019farsidemag,jeong2020farsidemag} and the generation of high-resolution pseudo-magnetograms from multi-wavelength observations \cite{park2019uvgeneration,shin2020pseudomag}. Instrumental advances further facilitate cross-observatory data harmonization \cite{jarolim2024instranslation}, while architectures such as convolutional neural networks (CNNs) and long-short-term memory (LSTM) models have demonstrably surpassed traditional statistical approaches \cite{bobra2015sharpsvm,huang2018deep,wang2020pilp,li2022knowledge,wang2022precursor,liu2023upsample,li2023dnncomplexar,deng2023twostage,wang2023video}. 

Recently, large artificial intelligence (AI) models have gradually been applied to the professional fields of Earth sciences and atmospheric sciences. Reichstein et al. \cite{reichstein2019earthsystem} pioneered process-informed deep learning for Earth system modeling, integrating physical constraints with data-driven approaches to enhance interpretability. Pangu-Weather \cite{bi2023accurate} uses a 3D neural network architecture to achieve high-resolution mid-range forecasts, significantly improving accuracy through hierarchical spatial modeling. GraphCast \cite{lam2023globalweather} uses graph neural networks (GNNs) for skillful medium-range predictions, which excel at capturing atmospheric dynamics via message passing across grid edges. Meanwhile, FengWu \cite{chen2023fengwuglobal} combines multi-modal data assimilation with a transformer backbone, pushing skillful global forecasts beyond 10 days through iterative error correction. FuXi \cite{chen2023fuxiglobal} adopts a cascade ML framework to extend high accuracy predictions to 15 days, using ensemble techniques for uncertainty quantification. XiHe \cite{Wang2024}, a data-driven global ocean eddy resolution forecasting model, uses a hierarchical transformer-based framework coupled with a land-ocean mask mechanism and an ocean-specific block, and achieves better forecast performance than existing leading operational numerical Global Ocean Forecasting Systems. GeoGPT \cite{ZHANG2024103976} combines large language models (LLM) and Geographical Information System (GIS) tools to autonomously handle geospatial tasks through natural language and help GIS professionals, with its "foundational plus professional" paradigm offering an effective approach for next-generation GIS development. In the case of the super solar storm in May 2024, AI tools \cite{Guastavino2025} also indicate the potential to identify solar situations and predict events: classify the morphological evolution of active regions through Vision Transformer (ViT), predict strong flares ($\geq$M-class based on CNN-LSTM video deep learning, and issue a one-hour advance warning of geomagnetic storms through the LSTM model based on in-situ data. This verified that the performance of AI in active region characterization, flare prediction, and geomagnetic storm warning is superior to traditional methods. Parallel innovations in other domains highlight potential breakthroughs in space weather.

Space weather forecasting operational agencies, through the research-to-operation (R2O) approach, have integrated some research-oriented methods such as solar feature recognition, active region magnetic field characteristic calculation, and flare prediction modeling into analysis and forecasting tools. These tools can provide references when forecasters analyze solar activity conditions and issue solar eruption forecast products. The Space Weather Prediction Center (SWPC)\footnote{\url{https://www.swpc.noaa.gov/}} from the National Oceanic and Atmospheric Administration (NOAA) and the Space Environment Prediction Center (SEPC)\footnote{\url{https://www.sepc.ac.cn/}} from the National Space Science Center, Chinese Academy of Sciences (NSSC, CAS) provide daily forecast products (e.g. alerts, watches and warnings) for solar activity. Hughes et al. \cite{2019JSWSC...9A..38H} adopted pixel-based machine learning algorithms to generate real-time solar thematic maps for SWPC operations, using a multi-channel data cube from EUV images and H$\alpha$ images to classify each spatial pixel, with expert consensus-labeled datasets as training/testing basis. It classifies eight solar features (quiet sun, bright region, etc.), contributing to timely space weather forecasting and guiding future algorithms by highlighting the need for spatial structure integration. The National Institute of Information and Communications Technology (NICT)\footnote{\url{https://swc.nict.go.jp/}} employs the Deep Flare Net model for operational predictions \cite{nishizuka2018deepflarenet}. The Community Coordinated Modeling Center (CCMC) hosts the Flare Scoreboard\footnote{\url{https://ccmc.gsfc.nasa.gov/scoreboards/flare/}}, an open platform that aggregates real-time forecasts from multiple institutions and conducts systematic assessments, offering critical benchmarks to improve operational systems.

Despite these advancements, persistent limitations remain. Sparse space weather measurements impede robust feature extraction, whereas inflexible architectures hinder the rapid integration of the eruptive magnetic topology dynamics \cite{amari2018magrope}. Currently, space weather forecasting models have yet to fully utilize multi-modal solar data, and expert experience has not established a positive cycle for the iterative enhancement of model capabilities. The processes of solar situation perception, analysis relying on massive data, and forecast production within space weather operational forecasting agencies are not fully integrated, and they still predominantly depend on human forecasters under the Observation-Orientation-Decision-Action (OODA) paradigm.

To address these gaps, we present a ``Solar Activity AI Forecaster''(SA-AI forecaster) - a scalable framework that dynamically integrates multi-modal observations through semi-supervised learning and human-in-the-loop validation while enabling continuous capability expansion with expert knowledge embedding through a modular architecture (Section~\ref{sec:AIf}). Our results (Section~\ref{sec:res}) achieve state-of-the-art forecast accuracy, demonstrating a viable path toward autonomous space weather forecasting.

\section{Solar Activity AI Forecaster}
\label{sec:AIf}

\subsection{Framework and Capability}

AI forecaster is designed to autonomously perform the tasks of a human forecaster in space weather forecasting agencies, including integrating and understanding the monitoring data, performing in-depth analysis of solar activity conditions, and providing forecasts.

\begin{figure}[h]
\centering
\includegraphics[width=0.9\textwidth]{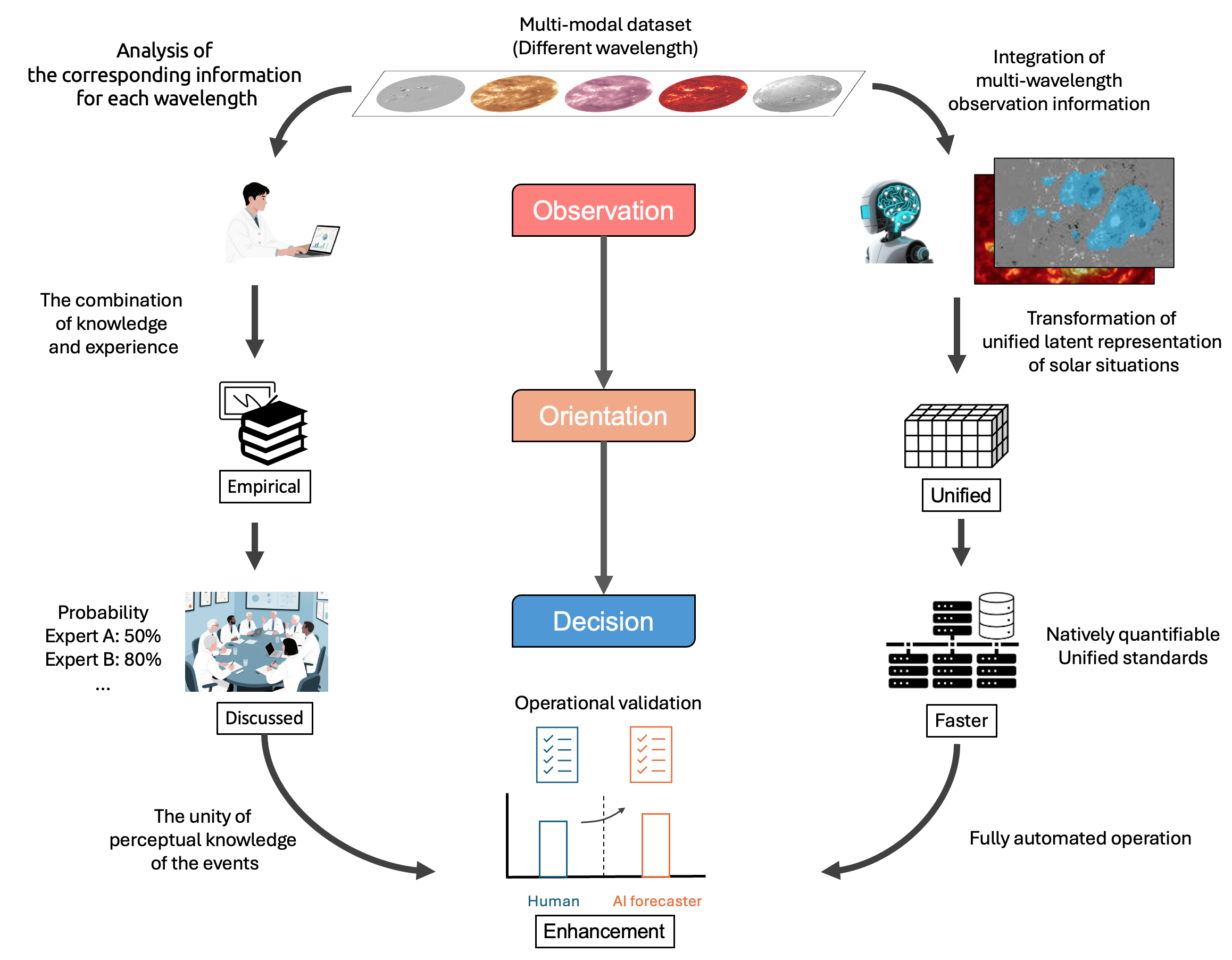}
\caption{Schematic diagram of the design of AI forecaster with comparison to the forecasting process of human forecasters under the Observation-Orientation-Decision-Action paradigm.}
\label{fig:design}
\end{figure}

Fig.~\ref{fig:design} is a visual representation illustrating the comparative aspects of human forecaster and AI forecaster under the Observation-Orientation-Decision-Action (OODA) paradigm. At first, human forecasters examine a variety of monitoring data from multiple satellites and numerous stations and establish a perceptual understanding of observational data. Then, human forecasters analyze environmental conditions based on their accumulated experience and knowledge. They also use software tools to calculate important physical parameters and precursor factors and make rational and intuitive assessments and summaries of environmental conditions. In this process, the expert knowledge and accumulated experience of the individual forecaster plays a dominant role. Finally, considering that individual interpretations and understandings may vary, they arrive at the most reliable forecasting results through group consultations and other collaborative methods.

The AI forecaster is implemented in the OODA paradigm using multiple models and modules. It adopts a dual-driven data-and-model architecture to independently accomplish the work of human forecasters, with each step being quantifiable and visualizable. At first, relying on computer hardware and large-scale model algorithms, it rapidly integrates and analyzes vast amounts of multi-source and multi-type data, and quantifies environmental conditions into standardized forms of representation. During the construction of solar activity situation perception models and solar eruption forecasting models, expert knowledge and experience are incorporated to provide more reliable and assessable, quantifiable forecasting results. Expert knowledge is the knowledge possessed by human experts that have been validated as effective for the task through human experience, statistical laws, etc. In this study, expert knowledge (priori knowledge, human experience, and annotation) is present at every stage of modeling training database preparation, task goal setting, and model selection of the SA-AI forecaster.

We have adopted the same strategy as the operational practices of space weather forecasting agencies, allowing AI forecaster and human forecasters to provide forecasts under the same operational conditions, thereby offering an objective assessment of their forecasting capabilities.

\begin{figure}[h]
\centering
\includegraphics[width=0.9\textwidth]{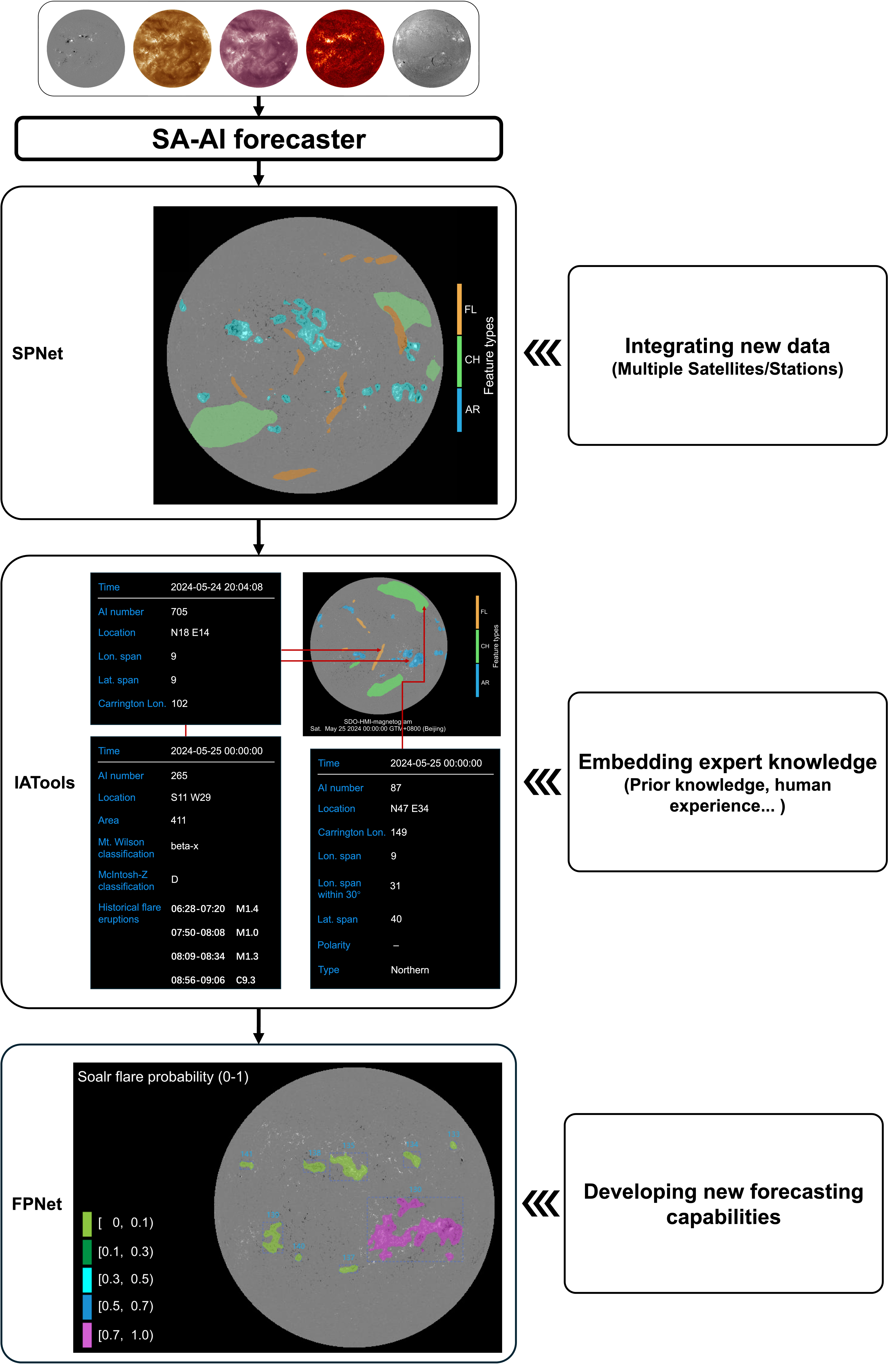}
\caption{Schematic diagram of the SA-AI forecaster framework and the visual output of each module. It includes the Situational Perception Module (SPNet), In-Depth Analysis Tools (IATools), and the Flare Prediction Module (FPNet).}
\label{fig:framework}
\end{figure}

The functional framework of the solar activity (SA) AI forecaster is shown in Fig.~\ref{fig:framework}. The SA-AI forecaster is made up of three modules: the Situational Perception Module (SPNet), the In-Depth Analysis Tools (IATools), and the Flare Prediction Module (FPNet). Through SPNet, AI forecaster first integrates a large amount of solar monitoring data to automatically generate daily solar situational awareness maps, converting important solar features (active regions (ARs), coronal holes (CHs), and filaments (FLs)) into data representations. Through IATools, the AI forecaster then independently integrates and summarizes real-time solar eruption information, using a variety of tools and incorporating expert knowledge to calculate physical feature parameters and conducting in-depth analysis and characterization of the solar activity conditions. Finally, through FPNet, the AI forecaster uses solar monitoring data and physical characterization of solar activity to provide probability forecasts of strong solar flares ($\ge$ M-class flares) for both the full-disk and active regions.

The SA-AI forecaster is scalable in three aspects: First, the SA-AI forecaster demonstrates strong generalization capabilities when dealing with multi-source data. In the future, more types of data can be incorporated from multiple satellites and various stations, thereby further enhancing the ability to perceive the environment. Second, during the construction of situation perception and forecasting models, the SA-AI forecaster can integrate more expert knowledge (important physical parameters and precursor factors). Through semi-supervised learning and a "human-in-the-loop" approach, it can continuously learn from the human forecasters' experience. Third, the SA-AI forecaster can improve its forecasting capabilities by developing and integrating new forecasting models, increasing the range of forecasting products, expanding the scope of forecasts, and improving the overall efficiency of forecasting.

Table~\ref{tab:functions} introduces the inputs, functions and the time required for fully automated operational processes of the three modules. The time required for the AI forecaster to complete one cycle of perception, analysis, and forecasting on a single NVIDIA Tesla V100S GPU is approximately 3 to 6 minutes.

\begin{sidewaystable}[ht]
\centering
\caption{Functions and operational runtime of the modular SA-AI forecaster}
\begin{tabular*}{\textwidth}{@{\extracolsep\fill}p{1cm}p{5.5cm}p{4.5cm}p{2cm}}
\toprule
\textbf{Module} & \textbf{Input Data} & \textbf{Functions Achieved/Results} & \textbf{Operational Runtime} \\
\midrule
\midrule
\multirow{2}{*}{SPNet} & 
Multi-modal solar images (including magnetograms, coronal EUV, H$\alpha$) &
Daily solar situational awareness map (including ARs, CHs, FLs) &
$\leq$2min \\
\addlinespace
\midrule
\multirow{3}{*}{IATools} &
Multi-modal solar images, X-ray flux observations and the results obtained from the SPNet module &
Physical characteristic parameters and summary of real-time flare information &
$\leq$2min \\
\addlinespace
\midrule
FPNet &
Various solar images, F10.7 flux, and the results obtained from the first two modules &
Strong flare ($\geq$ M-class) prediction for full-disk and active regions &
$\leq$2min \\
\botrule
\end{tabular*}
\label{tab:functions}
\end{sidewaystable}

\subsection{Data and Preparation}

The data used in this study span from January 1996 to May 2024 and include full-disk photospheric magnetograms, chromospheric H$\alpha$ images, extreme ultraviolet (EUV) coronal images, soft X-ray (SXR) flux measurements, and solar radio flux at 10.7 cm (2800 MHz). Photospheric magnetograms were obtained from SOHO/MDI \cite{Domingo2009soho,Scherrer1995mdi}, SDO/HMI \cite{Pesnell2012sdo,schou2012sdohmi}, and ASO-S/FMG \cite{gan2023asos,deng2019asosfmg}. Coronal EUV images were acquired by SOHO/EIT \cite{Domingo2009soho,delaboudini1995sohoeit}, SDO/AIA \cite{Pesnell2012sdo,lemen2012sdoaia}, and GOES-16/SUVI \cite{dranel2022goesrsuvi}, covering multiple spectral lines, including 171\text{\AA} (Fe,{\sc ix}), 193\text{\AA} (Fe,{\sc xii}, {\sc xxiv}), 195\text{\AA} (Fe,{\sc xii}), 211\text{\AA} (Fe,{\sc xiv}), and 304\text{\AA} (He,{\sc ii}). Chromospheric H$\alpha$ observations were provided by ground-based telescopes at BBSO, KSO, and HSOS, while SXR flux data were sourced from the GOES satellite series. The F10.7 observations were sourced from the Solar Radio Monitoring Program from Canada. All data were retrieved from publicly available repositories (see \textbf{Data Availability} for details). The list of historical outburst flares and the AR numbers are from NOAA. The SWPC publishes solar region summary (SRS) with AR numbers and solar event reports with flare events. Meanwhile, both SWPC and SEPC have real-time flare event monitoring report products to monitor flares erupting on the solar surface and issue timely warnings.

\begin{figure}[htbp]
\centering
\includegraphics[width=0.9\textwidth]{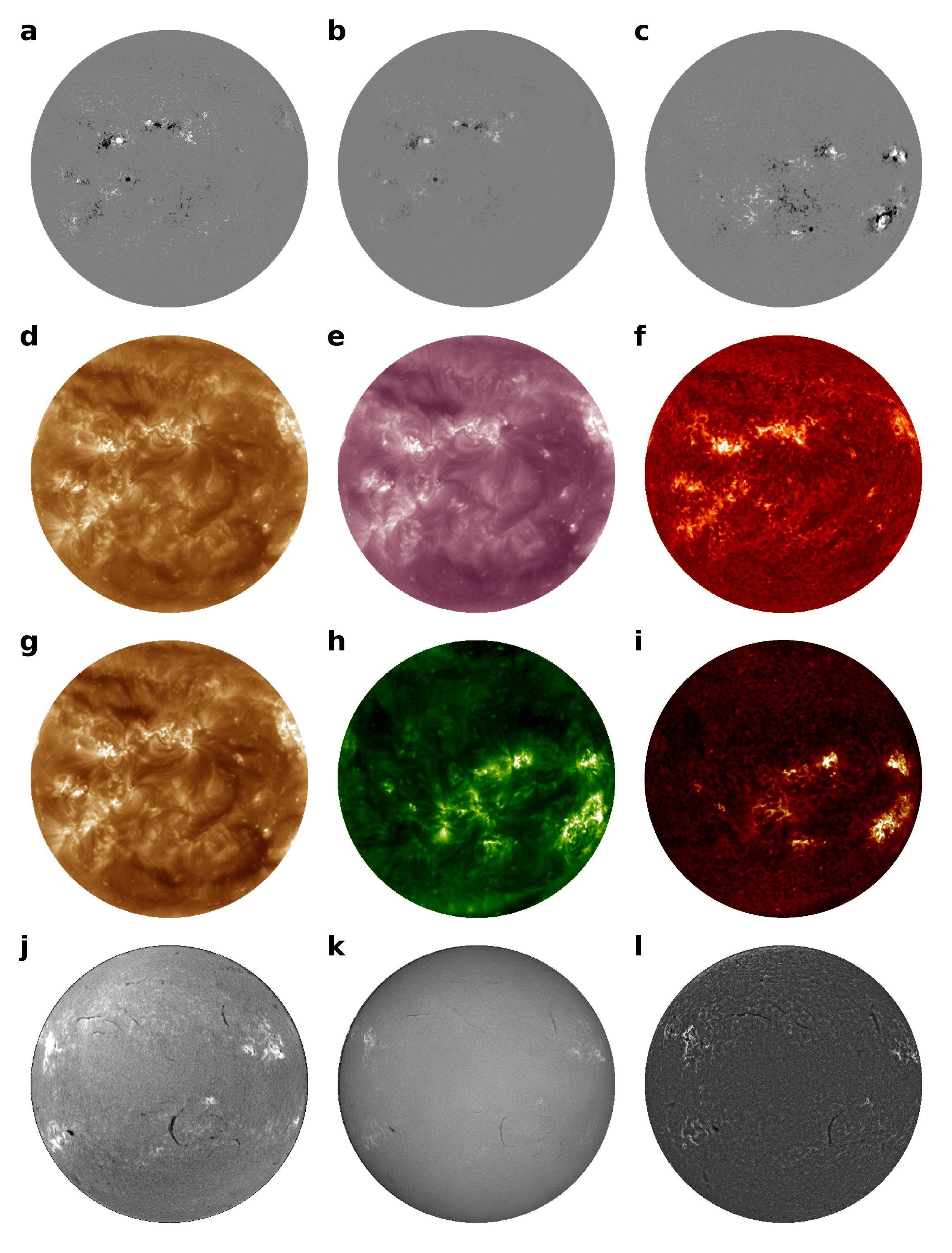}
\caption{Standard solar images after the preparation pipeline covering 1.1 solar radii. Panels (a)-(i) show standard images processed from SDO/HMI magnetogram at 00:00 UT, 2023 May 07, ASO-S/FMG magnetogram at 00:00 UT, 2023 May 07, SOHO/MDI magnetogram at 00:00 UT, 2003 November 02, SDO/AIA 193\text{\AA} at 00:00 UT, 2023 May 07, SDO/AIA 211\text{\AA} at 00:00 UT, 2023 May 07, SDO/AIA 304\text{\AA} at 00:00 UT, 2023 May 07, GOES/SUVI 195\text{\AA} at 00:00 UT, 2023 May 07, SOHO/EIT 195\text{\AA} at 01:19 UT, 2003 November 02, SOHO/EIT 304\text{\AA} at 00:00 UT, 2003 November 02, BBSO/H$\alpha$ at 00:00 UT, 2023 May 19, KSO/H$\alpha$ at 00:00 UT, 2023 May 20, and HSOS/H$\alpha$ at 00:00 UT, 2023 May 21 observations, respectively.}
\label{fig:standardpic}
\end{figure}

Data preprocessing followed a standardized pipeline. Photospheric magnetograms and coronal images, initially in FITS format, were converted to JPG format. Using FITS header metadata, we aligned the solar disk center with the image center and rescaled or cropped the images based on the solar radius. This yielded final images covering 1.1 solar radii with a uniform resolution of 512 × 512 pixels (Fig.~\ref{fig:standardpic}(a)-(i)). Similarly, H$\alpha$ images in JPG format were processed using the Hough transform to detect the center and radius of the solar disk, followed by rescaling to match the same spatial coverage and resolution (Fig.~\ref{fig:standardpic}(j)-(l)). These preprocessing steps ensure data consistency and reduce interference from solar rotation and projection effects.

\subsection{Methodology}

Current space weather forecasting models face critical limitations: over-reliance on single-modal data, inability to capture deep cross-modal correlations, and restricted analysis scope due to projection effects. Large models, particularly those based on multi-modal transformers, address these issues inherently. Their architecture enables dynamic integration of diverse data (e.g. magnetograms, EUV, and H$\alpha$ images) through cross-attention mechanisms, facilitating the learning of complementary information across modalities. Moreover, their global modeling capability overcomes projection-induced restrictions, supporting full-disk analysis essential for operational forecasting. This makes large models, such as Vision Transformers (ViT)~\cite{dosovitskiy2020image}, Masked Autoencoder (MAE)~\cite{he2022masked}, and Multi-modal MAE (MulMAE)~\cite{bachmann2022multimae}, viable solutions to bridge existing gaps.

SPNet and FPNet jointly serve as the foundation of the operational space weather large model and play a major role in the entire SA-AI forecaster. Their multi-modal design is fundamentally driven by the nature of solar activity. Key phenomena such as active region evolution and solar flares manifest through interdependent physical processes observed across distinct data domains. No single observation provides a complete physical picture. SPNet requires multi-modal fusion for holistic full-disk monitoring, detecting subtle precursors and tracking complexity, and FPNet's flare prediction necessitates integrating complementary data streams, for example, fusing historical flare indices with multi-wavelength solar observational images. This multi-modal approach is essential for capturing the intrinsic coupling of underlying physics, overcoming limitations of individual data sources, and generating the unified, physically-grounded representations required for accurate operational space weather forecasting.

\begin{figure}[htbp]
\centering
\includegraphics[width=0.9\textwidth]{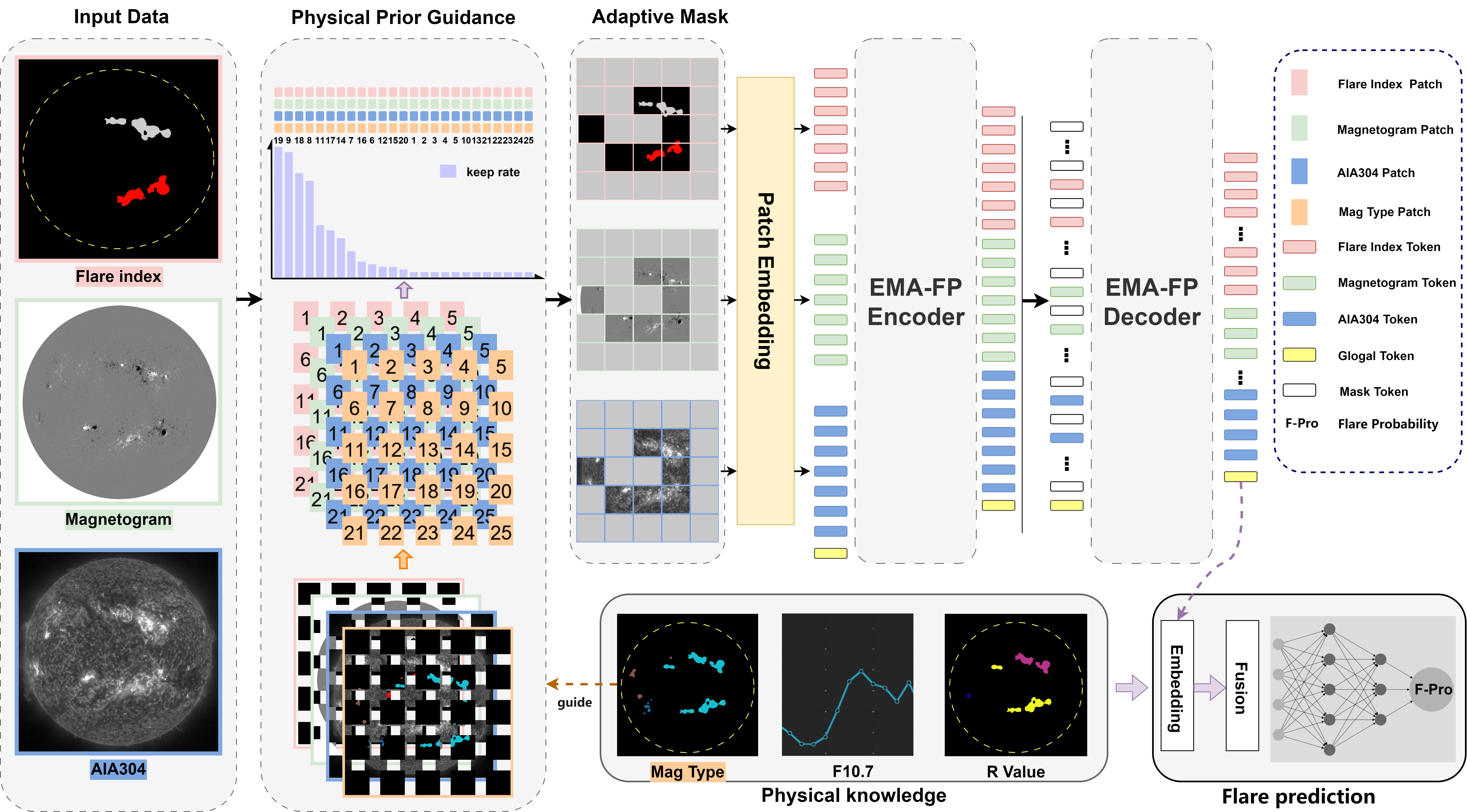}
\caption{Schematic overview of FPNet. The algorithm has three input branches for three types of inputs: AIA304\text{\AA} (bottom left), magnetograms (middle left) and flare index (top left). All input data are first processed by the Physical Prior-guided Adaptive Mask (PPAM) module, where the Physical Prior Guidance  module identifies activity-sensitive patches relevant to solar flares, and the Adaptive Mask module dynamically discards non-active patches while selectively preserving features critical to solar flares phenomena. These masked data are then fed into a transformer framework for further analysis. An additional global token is incorporated into the framework to learn deep associations across all input data. To enhance model performance by incorporating physical priors, the algorithm additionally utilizes several physical parameters such as Hale magnetic type, F10.7 and Magnetic neutral line R value. These physical parameters are first passed through a feature extractor to generate high-dimensional embeddings, which are then fused with the previously obtained global token. Finally, a fully connected layer is used to predict the probability of solar flares.}
\label{fig:aimod}
\end{figure}

SPNet and FPNet share the same core transformer architecture, jointly powering the SA-AI forecaster for operational space weather prediction. To specifically address the complex physics of solar flare eruptions, FPNet uniquely integrates a domain-informed physical knowledge module that extracts critical features from SPNet and IAtool data. Crucially, FPNet further incorporates a Physical Prior-guided Adaptive Masking (PPAM) module, which leverages solar physics knowledge to selectively preserve tokens by using probability-based asymmetric sampling technique within critical regions (e.g., active regions, flare precursors) in the input images. This targeted masking focuses the model’s reconstruction and representation capacity on the most salient features driving flares. These physics-refined image representations, along with the extracted physical features, are then fused within FPNet’s dedicated flare prediction network for enhanced forecasting. We detail the structure of FPNet as the representative framework, given its enhanced capability in modeling eruptive processes. Within our FPNet architecture (Fig.~\ref{fig:aimod}), we introduce an Efficient Masked Autoencoder for Flare Prediction (EMA-FP), specifically a hybrid framework that integrates heterogeneous solar data. The architecture processes spatiotemporally co-aligned imaging inputs, including AIA 304\text{\AA} observations, line-of-sight magnetograms, and flare history indices, through a multi-modal transformer encoder. Magnetogram-derived segmentation masks selectively preserve AR patches while masking quiescent regions, thereby generating multi-channel input patches. These inputs are then processed through a transformer self-attention mechanism to extract spatially distributed precursors. Concurrently, a physics feature mapper encodes scalar parameters (F10.7 cm radio flux, McIntosh active-region classifications) into high-dimensional embeddings aligned with the transformer's latent space. Nonlinear interactions are modeled through joint embedding layers, capturing eruption-correlated relationships. Both streams concatenate into a joint feature vector, which is transformed by a multi-layer perceptron (MLP) into probabilistic forecasts through a sigmoid output.

To adapt to the solar extremely dynamic range and rapidly evolving structures, which are fundamentally distinct from natural images, our FPNet develops a physics-aware masked reconstruction within its EMA-FP framework for solar-optimized pretraining. Specifically, our model consists of a PPAM mechanism, which decomposes into two core submodules: the Physical Prior Guidance module locates activity-sensitive regions, and the Adaptive Mask module generates a spatially aware mask that selectively preserves features within these identified active regions, a patch embedding module, a transformer encoder for feature extraction, and a task-specific decoder head for solar flare prediction. Building on MAE and MulMAE , our approach addresses the distributional disparities between natural images and solar data. Crucially, the PPAM replaces the random masking by generating magnetogram-derived masks that selectively preserve flare-productive active regions. This PPAM-enabled targeted masking enhances computational efficiency while maintaining the spatial context essential for precursor identification. By pretraining on the masked reconstruction of solar observations guided by PPAM, our model mitigates cross-domain transfer limitations, aligning learned feature representations with radiative and magnetic signatures indicative of eruptive behavior. The PPAM-generated masked multivariate solar observations are then processed by patch embedding, which partitions the input data into patches and projects each patch into a high-dimensional token embedding space. To preserve the inherent spatial structure of the solar data, two-dimensional sinusoidal positional embeddings are incorporated into these token embeddings. In addition, we further introduce a learnable global token to aggregate cross-modal information across the transformer layers, serving both as a hierarchical summarization of the evolving features and as the primary prediction anchor for the solar flare occurrence forecasting. These token embeddings are then fed into the transformer encoder, which consists of $L$ identical layers (where $L$ is a configurable hyperparameter), each comprising multi-head self-attention and multilayer perceptron blocks. Layer normalization is applied before each block and residual connections are applied afterwards. The encoder's final layer outputs are fed into a task-specific decoder for solar-flare prediction. This decoder is designed to learn task-relevant associations guided by the global token's holistic representation. Ultimately, the final state of this global token is fused with physical knowledge to predict the probability of solar flare occurrence.

By fusing multi-spectral images with quantitative physical parameters, EMA-FP establishes a unified representation of solar activity evolution. This holistic approach enhances the interpretability of precursor signals and demonstrates robust generalization across diverse solar conditions. Our study establishes domain-adapted transformers as a promising paradigm for advancing space weather prediction.

\section{Modeling and Results of SA-AI forecaster}
\label{sec:res}

\subsection{Situational Perception Module (SPNet)}
\label{sec:percep}

Active regions (ARs), filaments (FLs), and coronal holes (CHs) constitute the three primary solar features critical for space weather forecasting. ARs and FLs are potential sources of coronal mass ejections (CMEs). CHs drive coronal hole high-speed streams. ARs additionally serve as key indicators for solar flares and solar particle events (SPEs).

Situational Perception Module (SPNet) is a foundational model and also the first module of the SA-AI forecaster. It is designed to automatically integrate multi-modal solar images, identify the three important solar physical features (ARs, CHs and FLs), and generate daily solar situational awareness maps as shown in Fig.~\ref{fig:framework}.

To address the limitations of traditional models that rely on single-modal data, SPNet adopts a dual data-model driven framework. This framework dynamically integrates multi-type observations (e.g. magnetograms, EUV images, H$\alpha$ images) through multi-modal transformers, enabling the model to capture deep cross-modal correlations that simple CNNs cannot achieve. For example, by masking partial modal data (e.g., occluding H$\alpha$ images), the model is forced to reconstruct features using other modalities, thereby learning complementary information across different data types.

Initially, a data set was labeled by SEPC human forecasters for training the SPNet model, and the number of this labeled data set was limited. To overcome the scarcity of labeled solar features, SPNet employs a semi-supervised learning approach combined with human-in-the-loop validation. The process involves: 1) training the model using limited manually labeled data; 2) automatically labeling extended datasets through model inference; 3) expert correction of mislabeled samples; 4) retraining the model with corrected data. Ultimately, SPNet was constructed on the basis of a larger number of high-quality labeled datasets from 1996 to 2022. In addition, we prepared an independent dataset from January 2023 to May 2024 manually labeled by human forecasters, which was not involved in the model's training and iterative process, but was used independently for model testing (see Table~\ref{tab:labels}).

\subsubsection{Initial annotation challenges}

The annotation of active regions (ARs) is based on SDO/HMI images, the annotation of coronal holes (CHs) depends on SDO/AIA 193\text{\AA} images, and the annotation of filaments (FLs) is based on HSOS/KSO/BBSO H$\alpha$ images. Using LabelMe \footnote{\url{https://jameslahm.github.io/labelme/}}, human forecasters manually labeled ARs for 4070 days, CHs for 3311 days, and FLs for 1746 days, selecting one image per day from the data set that spans from 2010 to 2022 (see Table~\ref{tab:labels}).

The annotation process referred to some historical public data. Specifically, the annotation of ARs referred to the historical numbering of ARs by NOAA and the SHARP data set. When a SHARP patch contains multiple NOAA-numbered ARs, or when a single NOAA-numbered AR appears in multiple SHARP patches, we reannotated each active region to ensure that each patch in our annotated dataset corresponds one-to-one with a NOAA AR. The annotation of CHs and FLs referred to the daily solar synoptic map products \footnote{\url{https://www.swpc.noaa.gov/products/solar-synoptic-map}} from SWPC.

The annotation work is labor-intensive, so different time periods were assigned to different forecasters for labeling, which means that the annotations were subject to subjective influences. This initial manually annotated data set (Labels V1) is not comprehensive, as it omits ARs that were not numbered by NOAA, some polar CHs, and some FLs with smaller areas.

\subsubsection{Iterative training through the human-in-the-loop approach}

Two types of network were developed for SPNet: a single-modal model version that uses single-modality images as input and a multi-modal model that uses multiple types of images as input for the recognition of solar features. The recognition performance of solar features of the models was evaluated using the Intersection Over Union (IoU) metric (range: 0 to 100), which quantifies the accuracy of the label overlap. The higher mean value and lower standard deviation of the IoU indicate improved robustness.

Adopting MAE and multi-modal transformers similar to Fig.~\ref{fig:aimod}, we first trained the three single-modal model versions of SPNet (denoted SV1–SV3 in Table~\ref{tab:labels}) based on different training data sets. During model training, the foreground-background weight ratio significantly impacts the recognition area of ARs, CHs, and FLs: a higher ratio expands the identified solar feature areas, while a lower ratio contracts them. We tested different weight ratios (e.g. 16:1, 8:1, 6:1, 4:1, 2:1) on the validation set and selected those that maximized IoU for each feature: 6:1 for AR, 4:1 for CH and 2:1 for FL.

For the first training round, the model SV1 is trained in the initial manually annotated data set (Labels V1). Then, we had the model SV1 to re-label the data from 2010 to 2022. Compared to the initial manually annotated dataset, the model SV1 identified some ARs that were not numbered by NOAA, some polar CHs, and small FLs that had been overlooked. These correctly labeled solar features were added to the manually annotated data set, creating a new data set that combined human and machine annotations (Labels V2).

This new data set (Labels V2) was used to iteratively train the model SV2. Then, we had the model SV2 to label the ARs on SOHO/MDI images, the CHs on SOHO/EUV 195\text{\AA} images, and the FLs on HSOS/KSO/BBSO H$\alpha$ images from 1996 to 2009. Human forecasters reviewed all annotations provided by the model SV2, retained all the correct annotations, and removed the incorrect ones. This process generated an expanded data set from 1996 to 2022 that combined human and machine annotations (Labels V3), which was then used to iteratively train the model SV3. During this process, human experts corrected model errors and supplemented missing features (e.g., resolving 2394 SHARP patch ambiguities where multiple ARs numbered by NOAA shared a single patch) and labeled 4128 ARs since 1996 in total.

Compared to the initial annotated data set (Labels V1), we expanded the new annotated data set (Labels V3) to 6426 days (increased by 57\%) for AR, 10107 days (increased by 305\%) for CH, 3365 days (increased by 92\%) for FL.

\subsubsection{Evaluation of Solar Feature Recognition Capability}

In Table~\ref{tab:labels}), by comparing the IoU and Acc metrics for ARs and CHs of the three versions (SV1, SV2 and SV3), it can be seen that the recognition accuracy of the SPNet is gradually increasing. This iterative mechanism increased the number of solar feature labels, significantly improving the generalization of the model. Training in the expanded data set (Labels V3), the model SV3 achieved optimal generalization in the independent test set (mean IoU: 84.38 ± 2.57 for AR, 74.50 ± 9.45 for CH, and 72.00 ± 5.14 for FL; Acc: 92.29 ± 2.32 for AR, 90.15 ± 11.5 for CH, 87.57 ± 8.06 for FL in Table~\ref{tab:labels}). The improvement in Acc metrics is more significant than that of IoU. This is because IoU focuses more on the ratio of the intersection-union area between the prediction and the real sample, whereas the Acc evaluation is for each pixel. Therefore, the Acc metrics can better reflect that the improvement of labels by human-in-the-loop can enhance the model's pixel-level recognition accuracy for the solar features.

However, compared to SV1, the mean IoU and Acc for FLs of SV3 did not improve. One reason is that the model identified a relatively large number of small FLs in the independent test data set (January 2023- May 2024). These identifications were not errors. However, since these small dark stripes were not manually annotated, they lowered the values of the IoU and Acc evaluation metrics. This also demonstrates that the model not only accurately identified the larger dark stripes but also had a good ability to recognize the small dark stripes that were not sensitive in manual annotations.

In Fig.~\ref{fig:recg}, panels (a), (e) and (i) show the SDO/HMI magnetogram image with human annotation of ARs, SDO/AIA 193\text{\AA} image with human annotation of CHs, the BBSO H$\alpha$ image with human annotation of FLs, respectively. Panels (b), (f), and (j) show the version SV3 of SPNet's annotation of ARs, CHs, and FLs, respectively.

\subsubsection{Multi-modal Generalization}

It can be seen that the single-modal SPNet (SV3) is capable of effectively recognizing solar features from one type of image data. In practical operational scenarios, we may utilize data input from multiple satellite sources, and solar features can often be observed in various types of images. However, there are certain differences in observational characteristics, such as area and brightness. The annotation for the solar features in one type of solar image is not the same as in other types of solar image. In that case, if we were to re-enable the single-modal model to recognize solar features from other types of solar image, it would require a significant amount of additional labeling work. Instead, we adopted an alternative approach: training a multi-modal model to generalize the recognition capability of the current single-modal model to other solar images. Therefore, we trained a multi-modal model version of SPNet (denoted by MV1 in Table~\ref{tab:labels}), which is designed not only to extract solar features from multiple types of image data, but also to maintain the ability to recognize solar features from alternative image data when one type of image data is missing. The network that consists of the MAE and multi-modal transformers can also adopt multi-modal input, with one type of solar image defined as the primary modality and the other type defined as the secondary modality.

The multi-modal version (MV1) of the model demonstrated cross-wavelength capability:

(1) Identified ARs in SDO/HMI magnetogram images (primary modality) and 304\text{\AA} images (secondary modality).

(2) Recognized CHs in SDO/AIA 193\text{\AA} (primary modality) and 211\text{\AA} images (secondary modality).

(3) Identified FLs in HSOS/KSO/BBSO H$\alpha$ (primary modality) and SDO/AIA 304\text{\AA} images (secondary modality).

For example, the multi-modal verision (MV1) of the SPNet learns to relate the characteristics of CHs in two different images and can recognize ARs in either of them. Furthermore, training in the expanded data set (Labels V3), the MV1 model improved the mean IoU in the independent test set by 2.08\% for AR, 1.34\% for CH and 2.27\% for FL, Acc by 1.20\% for AR and 1.26\% for FL (in Table~\ref{tab:labels}), compared to the model SV3.

\begin{figure}[h]
\centering
\includegraphics[width=0.9\textwidth]{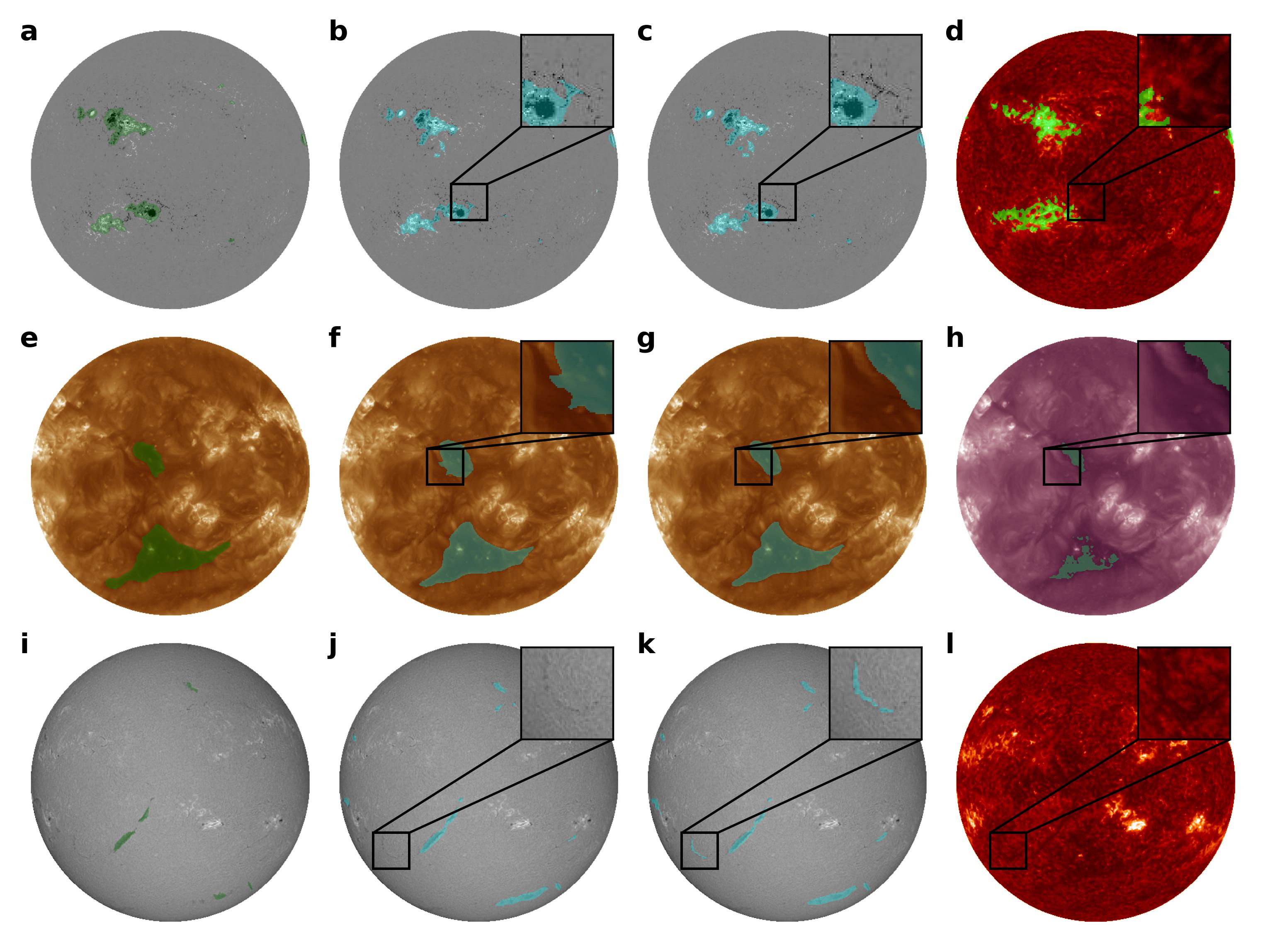}
\caption{Illustration of human experts' annotation and SPNet's annotation of solar features on solar images. The first column is the human experts' annotation. The second column is the single-modal SV3 of SPNet's results. The third column is the multi-modal MV1 of SPNet's results on the primary modality. The fourth column is the multi-modal MV1 of SPNet's results when removes the primary modality and only inputs the secondary modality.}
\label{fig:recg}
\end{figure}

In Fig.~\ref{fig:recg}, panels (c), (g), and (k) show the version MV1 of SPNet's annotation of ARs, CHs, and FLs when taking two types of solar images as input, respectively. It can be seen that in panel (l), a slender filament located at the eastern edge of the solar disk is well identified, which was not detected by the manual annotation (in panel (i)) and the single-modal SV3 annotation (in panel (j)). It can also be observed that even in the absence of the primary modality and with only the secondary modality as input, the MV1 model is still able to recognize solar features that are consistent with the characteristics of the input image itself. However, when only the AIA304\text{\AA} images are taken as input, the MV1 model cannot identify the filaments.

Unlike SWPC's pixel-to-pixel classification method \cite{2019JSWSC...9A..38H}, which lacks deep cross-modal learning, SPNet's multi-modal Transformer architecture enables ``global-local'' feature interaction. For example, SPNet can learn the morphological correspondence between ARs in magnetograms and AIA 304\text{\AA} images, dynamically expanding new feature types without relying on predefined categories. This significantly enhances the model's adaptability to different observation instruments and wavelengths.

The multi-modal SPNet is capable of identifying features in any individual image and capturing the actual characteristics of these features within the image. This ability allows the SA-AI forecaster to excel in handling incomplete data sets. For instance, even when data from certain modalities is missing, the model can still infer and complete the missing parts through information from other modalities.

\subsubsection{Cross-Instrument Validation}

The SPNet was generalized to similar data from other instruments. In Table~\ref{tab:labels}, it can be found that:

(1) Tested in the ASO-S/FMG magnetograms images from (January 2023 to May 2024, the SPNet exhibited a good performance generalization (mean IoU: 75.87 ± 4,33) for recognition of ARs.

(2) Tested in GOES-16/SUVI 193\text{\AA} images, despite different resolutions and sources, SPNet also demonstrated excellent performance (mean IoU: 74.15 ± 8.80) for recognition of CHs, which is close to the results obtained on SDO/AIA 193\text{\AA} images.

The training and testing sets for the FLs recognition models of SPNet already include H$\alpha$ images from multiple sources (HSOS/KSO/BBSO), which also demonstrate that the model has good generalization ability.

SPNet maintains high recognition accuracy across different observation instruments. When a certain modal data is missing (e.g., SDO/HMI magnetograms images), the model can infer through other modalities (e.g., ASO-S/FMG magnetograms images), demonstrating strong cross-instrument generalization.

\begin{sidewaystable}[ht]
\centering
\caption{Performance evaluation of SPNet on independent test set (January 2023-May 2024).}
\label{tab:labels}
\begin{tabular}{l l l l r r r r}
\hline
\textbf{Features} & \textbf{Model} & \textbf{Training Set} & \textbf{Testing Set} & \textbf{$IoU_{\text{Mean}}$} & \textbf{$IoU_{\text{STD}}$} & \textbf{$Acc_{\text{Mean}}$} & \textbf{$Acc_{\text{STD}}$} \\
\hline\hline
\multirow{5}{*}{ARs} & SV1 & HMI (2010-2022, 4070days) & HMI & 84.81 & 2.60 & 91.74 & 2.56 \\
 & SV2 & HMI (2010-2022, 4070days) & HMI & 84.94 & 2.67 & 91.75 & 2.62 \\
 & SV3 & MDI+HMI (1996-2022, 6426days) & HMI & 84.38 & 2.57 & 92.29 & 2.37 \\
 & SV3 & & ASO-S/FMG & 75.87 & 4.33 & 86.77 & 5.85 \\
 & MV1 & HMI+AIA 304{\AA} (2010-2022, 3787days) & HMI+AIA 304{\AA} & 86.14 & 2.51 & 93.49 & 2.37 \\
\midrule
\multirow{5}{*}{CHs} & SV1 & AIA 193{\AA} (2010-2022, 3311days) & AIA 193{\AA} & 74.70 & 9.37 & 86.76 & 12.19 \\
 & SV2 & AIA 193{\AA} (2010-2022, 3311days) & AIA 193{\AA} & 75.00 & 9.81 & 89.48 & 11.98 \\
 & SV3 & EUV 195{\AA}+AIA 193{\AA} (1996-2022, 10107days) & AIA 193{\AA} & 74.50 & 9.45 & 90.15 & 11.50 \\
 & SV3 & & GOES-16/SUVI 193{\AA} & 74.15 & 8.80 & 91.36 & 10.62 \\
 & MV1 & AIA 193{\AA}+211{\AA} (2010-2022, 3311days) & AIA 193{\AA}+211{\AA} & 75.50 & 9.80 & 89.51 & 11.67 \\
\midrule
\multirow{4}{*}{FLs} & SV1 & BBSO+HSOS+KSO H$\alpha$ (2010-2022, 1746days) & H$\alpha$ & 73.20 & 5.45 & 87.93 & 8.08 \\
 & SV2 & BBSO+HSOS+KSO H$\alpha$ (2010-2022, 1746days) & H$\alpha$ & 72.10 & 5.21 & 86.10 & 8.33 \\
 & SV3 & BBSO+HSOS+KSO H$\alpha$ (1996-2022, 3365days) & H$\alpha$ & 72.00 & 5.14 & 87.54 & 8.06 \\
 & MV1 & H$\alpha$+AIA 304{\AA} (2010-2022, 1746days) & H$\alpha$+AIA 304{\AA} & 73.64 & 5.68 & 88.80 & 7.65 \\
\hline
\end{tabular}
\end{sidewaystable}

\subsection{In-depth Analysis Tools (IATools)}
\label{sec:analysis}

In-depth Analysis Tools (IATools) is the second module of the SA-AI forecaster. It is a collection of toolboxes designed for in-depth analysis of the solar situational awareness maps obtained by the SPNet. It automatically provides various quantified physical parameters and several important descriptive definitions of solar features and summarizes historical outbreak information for active regions, as shown in Fig.~\ref{fig:framework}. These outputs are used for subsequent forecasting modeling of the solar eruptions.

Firstly, by IATools, the discrete points of the solar features initially recognized by the SPNet were identified as independent entities of active regions, coronal holes, and filaments through the connectivity and clustering approach.

For the identification of ARs, the Density-Based Spatial Clustering of Applications with Noise (DBSCAN) algorithm was employed. DBSCAN defines clusters as maximal sets of density-connected points, enabling the partitioning of sufficiently dense regions into distinct clusters while identifying arbitrary-shaped groupings within noisy spatial datasets. In contrast, CHs and FLs were identified using connectivity-based methods, where each isolated connected region was treated as an individual CH or filament. When the boundary distance between two adjacent CHs/FLs fell below a predefined threshold, the regions merged into a single entity. Subsequently, small-area features were filtered on the basis of the calculated areas of ARs, CHs, and FLs.

This pipeline achieved a robust identification of the three solar features and then applied numbering and tracking considering the regional overlap after removing the effects of the Sun’s overall rotation.

For the three solar features, the IATools was required to be able to automatically calculate various physical parameters, which are capable of characterizing the intrinsic physical properties of the features and have potential statistical correlations with solar eruptive activity (e.g. flares, CMEs) or solar wind variations.

\subsubsection{Active Region's Parameters Calculated by IATools}

For each AR, we compute the following parameters.

(1) Heliographic coordinates (central position) and Carrington longitude (for solar rotation tracking).

(2) The magnetic neutral line R value \cite{schrijver2007rvalue}, quantifying the magnetic shear energy.

(3) The Hale magnetic type \cite{Hale1919} and McIntosh morphological class \cite{McIntosh1990}, describing the complexity of AR.

(4) The Flare Index (FI) quantifying cumulative flaring activity ($\geq$ C5-class flares) during the preceding 24-hour period for each AR.

Statistical analysis revealed strong correlations between these parameters and flare productivity. In particular, ARs with high R values, $\delta$-class Hale types, or complex McIntosh classifications exhibited elevated flare rates.

By IATools, the calculation procedure of FI involves first utilizing GOES 1-8\text{\AA} soft X-ray flux measurements to identify flare events on the solar disk, classifying each detected flare according to its peak flux intensity and timestamping upon onset and upon end times. IATools obtains the near-real-time flare alert information in accordance with the rules of SEPC's alert products. Then, AIA 193\text{\AA} coronal imaging data were applied during flare occurrence intervals to precisely map the spatial origin of each flare event and correlated with the pre-identified active region to assign the active region number for each flare. Finally, the Flare Index is computed using the weighted sum of flare counts categorized by their GOES classes, defined as:

\begin{equation}
\text{FI} = 1 \times \sum\text{C} + 10 \times \sum\text{M} + 100 \times \sum\text{X},
\end{equation}

where $\sum\text{C}$, $\sum\text{M}$, and $\sum\text{X}$ denote the total number of C-class, M-class, and X-class flares detected within the active region over the preceding 24 hours, respectively.

IATools consists of two magnetic classification models that output AR for the Mt. Wilson magnetic classification and the McIntosh-Z classification, respectively. Here, it classifies the Mt. Wilson magnetic classification into three classes, which are alpha, beta, and beta-x (containing gamma, beta-gamma, delta, delta-gamma, beta-gamma-delta, beta-delta) classes; the McIntosh-Z classification into seven classes, which are A, B, C, D, E, F, and H.

In the training process, to address the longitude-dependent projection effects in solar ARs (significant distortion beyond ±30$^{\circ}$ longitude vs. mild distortion within ±30$^{\circ}$), we implements a unified training framework comprising two architecturally identical deep convolution neural networks (DCNNs) dedicated to distinct longitudinal zones. The training protocol incorporates geometric/photometric augmentation strategies (random 90-degree rotations, scaling, translation, and brightness perturbations) to mitigate class imbalance and enhance generalization. The main part of the network consists of a two-layer Conv-ReLU-MaxPool module constituting the feature extraction layer, which is spread and flattened by a 128-dimensional fully connected layer for higher-order semantic abstraction, and overfitting is suppressed by Dropout regularization at the end. Both models are trained end-to-end with the Adam optimizer to achieve accurate modeling of the AR Mt. Wilson magnetic classification and McIntosh-Z classification.

In the inference phase, for each AR, IATools dynamically selects a dedicated model based on the longitude of the geometric center of its Stonyhurst coordinate system (±30$^{\circ}$ outside/inside). The preprocessing is based on the minimum envelope square of 5\% outward expansion of the active area boundary, and 160×160 standardized inputs are generated by mean padding (based on the edges pixels of the active area); the normalized inputs are fed into the network, and the classification probability distributions are output through the Softmax layer to achieve millisecond-level real-time inference. This process effectively overcomes the interference of projected deformation with classification confidence through geometric normalization and robust feature extraction techniques.

\begin{figure}[h]
\centering
\includegraphics[width=0.8\textwidth]{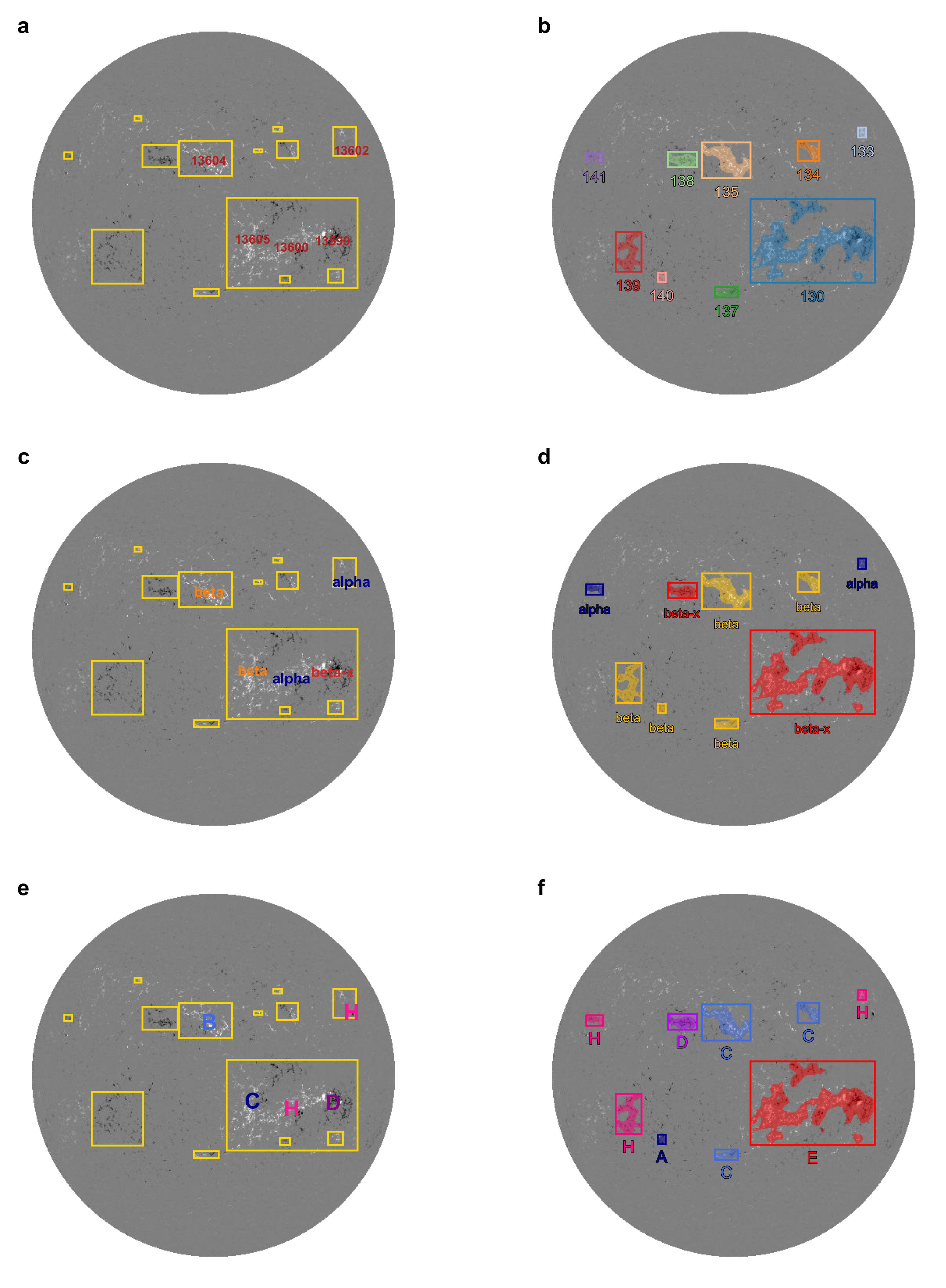}
\caption{Illustration of the identification and magnetic classification of Active regions by IATools on the magnetogram image observed by SDO/HMI at 00:00 UT, 2023 March 11. Panels (a), (c), and (e) show the five ARs numbered and classified by NOAA and the 13 SHARP AR patches at the same moment. Panels (b), (d), and (f) show the nine ARs numbered and classified by IATools.}
\label{fig:ARparameter}
\end{figure}

Fig.~\ref{fig:ARparameter} shows the magnetic classification results obtained by IATools in the right column. Panel (b) shows the nine ARs automatically numbered by IATools in the HMI magnetogram image. As a reference for comparison, panel (a) shows the identification bounding boxes (yellow) referring to the 13 SHARP AR patches and the corresponding five NOAA AR numbers (red) located at their center positions at the same moment. As can be seen in the two panels, AR 13604 and AR 13602 are identified by SHARP and IATools. However, because of the relatively close structure of AR 13599, AR 13600, and AR 13605 on the solar disk, both SHARP and IATools have merged these three active regions into one. Panels (d) and (e) show the Mt. Wilson magnetic type and the McIntosh-Z type classified by IATools, respectively. As a reference for comparison, panels (c) and (e) show the ARs' magnetic type classified by NOAA.

\subsubsection{Coronal Hole's Parameters Calculated by IATools}

For each CH, we compute the following parameters.

(1) Central heliographic coordinates and Carrington longitude (for solar wind source tracking).

(2) Magnetic polarity (unipolar dominance) and longitudinal span (within ±30° of disk center).

These parameters are critical for solar wind modeling, as CHs with broader longitudinal extents and persistent polarity correlate with high-speed solar wind streams at Earth.

\subsubsection{Filament's Parameters Calculated by IATools}

Filaments were characterized by:

(1) Heliographic coordinates for the central and endpoint, enabling tracking of their spatial evolution across the solar disk.

(2) Carrington longitude, linking their positions to the solar magnetic field structure.

Multiple positional points further allowed for analysis of filament migration patterns during the solar cycle, with polar filaments showing distinct latitudinal drift compared to low-latitude ones.

\subsubsection{Evaluation of IATools}
\label{sec:detailsset}

To verify the reliability of IATools, we performed quantitative evaluations using an independent test set (January 2023 to May 2024) with manual annotations from experts as ground truth.

\textbf{Performance of Entity Identification}

According to the SWPC's daily ``solar region summary'' products under operational conditions, there are a total of 3925 AR samples (involving 472 NOAA AR numbers) for a total of 511 days in this test period. Active regions without sunspots will not be recorded in SWPC's daily ``solar region summary'' products. However, IATools autonomously identified 5746 AR samples (involving 1006 SEPC AR numbers). The number of flare samples differs from that of SWPC as a result of the different ways of identifying, tracking of ARs, and localizing flare events.

The adjusted Rand index (ARI) is a normalized measure of agreement between the clustering results and the reference classifications, adjusted to account for chance agreements, ranging from -1 to 1. ARI close to 1 indicates nearly perfect alignment with the reference. For AR clustering (DBSCAN), the ARI and the IoU between IATools and manually annoations of SEPC’s experts reached 0.54 and 51.97, indicating great consistency in distinguishing overlapping ARs. Since there is a difference between the SPNet segmentation results and the manual entity division, the ARI and IoU calculated for clustering under this condition will cause the score to be low. AR features with small area were filtered as noise after DBSCAN, the ARI and IoU increased to 0.58 and 52.42, respectively. During this process, the small area feature regions filtered accounted for an average of 16\% of the total number of feature regions, while their area represented an average of 0.9\% of the total area of all feature regions.

\textbf{Evaluation of Computational Consistency of the Flare Index (FI)}

According to the real-time monitoring reports (``solar and geophysical event reports) by SWPC, there were a total of 258 days of strong flares (M-class or above flare), with a total of 691 strong flares involving 136 NOAA AR numbers and 28 strong flares without the corresponding NOAA AR numbers.

However, the identification of flare events and mapping them to source regions on the solar disk by IATools is based on GOES soft X-ray flux and AIA 193\text{\AA} coronal imaging data, the absence of which can result in some flare events not being recognized. For example, SWPC identified an M1.1 flare peak at 05:18 UT, 2023 February 13 that was not identified by IATools according to SEPC's flare alert rule (by SEPC, a flare is identified based on 1-minute resolution X-ray flux observation data, which requires four consecutive points showing a monotonic increase, with the fourth point being greater than or equal to 1.4 times the first point). In this time period, in accordance with the rules for flare real-time monitoring report products established by SEPC, IATools identified a total of 256 days with a total of 770 strong flares involving 128 SEPC AR numbers, and two strong flares without the corresponding SEPC AR numbers.

The Pearson correlation coefficients between the daily full-disk's accumulated FI calculated by IATools and the SWPC's records is 0.94. The Pearson correlation coefficient between the daily AR's maximum FI calculated by IATools and the SWPC's records was 0.93. It confirms its reliability in quantifying past flaring activity. This high accuracy not only validates the robustness of the FI calculation formula but also indirectly attests to the precision of two critical upstream processes in IATools: flare event identification and spatial localization to active regions. Specifically, FI computation relies on accurately detecting flare events (via GOES soft X-ray flux measurements) and precisely mapping each event to its originating active region (using AIA 193\text{\AA} images to pinpoint spatial origins). The strong consistency with SWPC’s records thus indicates that IATools effectively avoids misidentifying non-flare events, correctly classifies flare magnitudes, and reliably associates each flare with its corresponding active region, which are the key prerequisites for meaningful FI quantification and subsequent flare prediction.

\textbf{Magnetic Classification Performance}

The dual DCNNs model for Hale-type classification achieved an overall F1 score of 0.61 on the test set, with an F1 score of 0.44 for the $\delta$-class (critical for strong flares) and 0.69 for the $\beta$-class. This outperformed single DCNN models (0.54 overall F1 score) by mitigating projection effects in off-center ARs.

McIntosh classification showed a weighted F1 score of 0.40, with high precision (0.49) for complex classes (e.g., E, F) that are strongly associated with M-class flares.

\subsection{Flare Prediction Module (FPNet)}
\label{sec:forcasting}

Flare Prediction Module (FPNet) is a foundational model and also the third module of the SA-AI forecaster. It is constructed based on the data and the analysis results obtained from SPNet and IATools, building an end-to-end model as shown in Fig.~\ref{fig:aimod} to achieve the forecast of strong solar flare eruptions as shown in Fig.~\ref{fig:framework}.

FPNet predicts the probability forecast and binary forecast of M-class or stronger flares for the solar disk and each AR within the next 24 hours, by integrating the solar images (the full-disk photospheric magnetograms and chromospheric activity indicators-AIA 304$\text{\AA}$)), the physical parameters of ARs (AR magnetic type (Hale/McIntosh)), the current overall solar radiation condition (the F10.7 flux), and the contextual eruption history of ARs (flare index).

FPNet, leveraging multi-modal large model architectures, specifically resolves three key challenges in traditional flare prediction:

(1) Multi-modal integration inefficiency: FPNet leverages physic priors to reveal the inherent relationships among different solar imaging modalities, integrating multi-source data with physical parameters via a dedicated fusion mechanism.

(2) Limited full-disk prediction: FPNet captures long-range dependencies within individual full-disk observations and to model complex interactions across multi-modal full-disk data, enabling comprehensive solar flare forecasting across the entire solar disk.

(3) Imbalanced data handling: Using probability-based asymmetric sampling technique, FPNet up-weights flare-associated pixels while preserving global context, enhancing sensitivity to rare strong-flare events.

The evaluation of FPNet as an independent forecasting model is introduced in the next section. The evaluation of the SA-AI forecaster, which integrates three modules for fully automatic operation, will also be presented in the next section.

\subsubsection{Evaluation Methods of Flare Forecasting Capability}

FPNet provides both the probability forecast and the binary forecast for strong-flare in the next 24 hours.

For probabilistic solar flare forecasts, three key metrics are employed: the Brier Score (BS), Brier Skill Score (BSS), and the area under the Relative Operating Characteristic curve (ROCA). These metrics assess the accuracy, relative performance, and discriminative power of probabilistic predictions, respectively \cite{Wilks2006, Cui2016}.  

\textbf{Brier Score (BS)}

BS quantifies the mean squared error between predicted probabilities and observed binary outcomes (1 for the flaring case, 0 for the non-flaring case):  

\begin{equation}
\text{BS} = \frac{1}{N}\sum_{i=1}^{N}(y_i - \hat{y}_i)^2,
\end{equation}
 
where $N$ is the number of samples, $y_i$ is the observed value, and $\hat{y}_i$ is the predicted probability. A BS of 0 indicates perfect accuracy, while 1 reflects complete inaccuracy.  
 
\textbf{Brier Skill Score (BSS)}  

BSS evaluates the performance of the model relative to a non-skilled reference forecast (e.g., climatological averages):  

\begin{equation}
\text{BSS} = 1 - \frac{\text{BS}_{\text{forecast}}}{\text{BS}_{\text{reference}}}.
\end{equation}

A BSS of 1 denotes perfect skill, whereas negative values indicate poorer performance than the reference.  

\textbf{Relative Operating Characteristic (ROC) Curve and ROCA}

The ROC curve plots the recall (true positive rate) against the false alarm rate (false positive rate) across varying probability thresholds. The area under this curve (ROCA) ranges from 0 to 1, with 1 representing ideal discrimination. Points near the top-left corner of the ROC curve signify optimal performance, while those along the diagonal indicate random guessing.

\textbf{Binary Classification Metrics} 

Probabilistic forecasts can be converted to binary "YES/NO" predictions using a threshold. Performance is evaluated using a confusion matrix that includes the following four categories.

\textbf{True Positive (TP)}: The number of flaring samples successfully classified.

\textbf{True Negative (TN)}: The number of non-flaring samples correctly classified.

\textbf{False Positive (FP)}: The number of non-flaring samples wrongly classified as 'flaring'.

\textbf{False Negative (FN)}: The number of flaring samples wrongly classified as “non-flaring”.

Key derived metrics include:

\begin{equation}
\text{Precision} = \frac{\text{TP}}{\text{TP} + \text{FP}},
\end{equation}
\begin{equation}
\text{Recall} = \frac{\text{TP}}{\text{TP} + \text{FN}},
\end{equation}
\begin{equation}
\text{F1 Score} = \frac{2 \cdot \text{Precision} \cdot \text{Recall}}{\text{Precision} + \text{Recall}},
\end{equation}
\begin{equation}
\text{Accuracy} = \frac{\text{TP} + \text{TN}}{\text{TP} + \text{FP} + \text{FN} + \text{TN}},
\end{equation}
\begin{equation}
\text{CSI} = \frac{\text{TP}}{\text{TP} + \text{FP} + \text{FN}},
\end{equation}
\begin{equation}
\text{FAR} = \frac{\text{FP}}{\text{TP} + \text{FP}},
\end{equation}
\begin{equation}
\text{HSS} = \frac{2[(\text{TP} \cdot \text{TN}) - (\text{FN} \cdot \text{FP})]}{(\text{TP} + \text{FN})(\text{FN} + \text{TN}) + (\text{TP} + \text{FP})(\text{FP} + \text{TN})},
\end{equation}
\begin{equation}
\text{TSS} = \frac{\text{TP}}{\text{TP} + \text{FN}} - \frac{\text{FP}}{\text{FP} + \text{TN}}.
\end{equation}

In this study, we have calibrated the probabilistic forecasting model so that the binary classification model uses a threshold of 0.5 from the probabilistic forecasting model.

\textbf{Validation Against Reference Forecasts} 

Model performance is benchmarked against operational forecasts from institutions such as the SWPC of NOAA and the SEPC of NSSC, CAS. These centers provide daily probabilities of M- and X-class flares over 1–3 days. Furthermore, the CCMC’s Flare Scoreboard offers comparative results from models such as ASSA, NOAA. A baseline SVM model, incorporating solar active region parameters (e.g., McIntosh classification, 10 cm radio flux), is used for further validation. These forecast results can be found on the website of the Data Availability section.

\subsubsection{Evaluation of FPNet as an independent forecasting model}

We first evaluate FPNet as an independent model in a consistent test set (from January 2023 to May 2024). In this test, the input provided to FPNet was consistent with the information available to human forecasters, to evaluate the forecasting results that the FPNet model can produce. To distinguish it from the fully automatic SA-AI forecaster (consisting of SPNet, IATools and FPNet), it is labeled as FPNet* in the following evaluation section. In Table~\ref{tab:model_inputs}, the input data for FPNet* include the observational data (solar images and the F10.7 index), human annotation based on the ARs' information (NOAA number, location, area, magnetic type) from SWPC's daily reports, and the physical parameter R value and flux index calculated based on the labels.

We used a test sample dataset that is consistent with the operational running condition of SWPC according to the SWPC's daily ``solar region summary'' and ``solar and geophysical event reports'' products (see details in Section~\ref{sec:detailsset}). In this test set, both the binary classification and probabilistic prediction results of space weather agencies (SWPC and SEPC), the CCMC Flare scoreboard (NOAA, ASSA) and the baseline model (SVM) were evaluated. The results were listed in Table~\ref{tab:forecast_eval}. The radar chart in Fig.~\ref{fig:metrics} visualizes the performance of multiple models across seven key evaluation metrics: accuracy, precision, recall, F1, TSS, HSS, and CSI (normalized to a scale of 0-1 for consistency). In the radar chart, the closer each individual evaluation metric is to 1, the better the model performs; with balanced consideration of multiple metrics, a broader overall profile of the polygon indicates a superior model. Fig.~\ref{fig:roac} shows the probabilistic prediction ROC curves. In ROC curve analysis, a model is deemed better if it has a larger area under the curve (ROCA), reflecting a stronger ability to distinguish between positive and negative classes across all threshold settings.

We evaluated FPNet* using two complementary approaches: probabilistic prediction and binary classification. For probabilistic prediction, the direct output of the eruption probabilities (0-1 range) is used. For binary classification, a fixed threshold (0.5) is used for general comparison, as well as TSS-optimized thresholds (listed as "Th" in Table~\ref{tab:forecast_eval}) to account for uncalibrated forecasts (0.2 for AR's forecasts by SWPC, 0.3 for full-disk's forecast by NOAA, and 0.4 for full-disk forecast by ASSA). It should be noted that FPNet* only takes a fixed threshold (0.5).

By comparing the key metrics in Table~\ref{tab:forecast_eval}, it can be found that SWPC performs better than SVM and achieved an F1 score of 0.4286, a TSS score of 0.4549, a BSS of 0.1877, and a ROCA of 0.8145 for AR's flare prediction. Compared to others, FPNet* performed best with the highest F1 score of 0.5020, a TSS score of 0.4392, a BSS of 0.2409, and a ROCA of 0.9065 for AR's flare prediction. That is, FPNet* outperformed human forecasters for AR's flare prediction, surpassing SWPC by 17\% in the F1 score and by 11\% in the ROCA score. According to Fig.~\ref{fig:metrics}(b), FPNet* outperforms SWPC in precision, F1 and HSS, but falls slightly behind in recall, creating a wider overall profile than SWPC.

For the full-disk's flare prediction, SEPC achieved an F1 score of 0.7140, a TSS score of 0.3858, and a ROCA of 0.7450. SWPC achieved an F1 score of 0.6821, a TSS score of 0.4636, and a ROCA of 0.8231. Compared to others (SWPC, SEPC, NOAA, ASSA, and SVM), FPNet* performs best with the highest F1 score of 0.7183, a TSS score of 0.4442, and a ROCA of 0.7685. That is, FPNet* also outperformed human forecasters for the full-disk's flare prediction, surpassing SEPC by 1\% in the F1 score, 15\% in the TSS score, and 3\% in the ROCA. According to Fig.~\ref{fig:metrics}(a), SWPC exhibits a pronounced strength in precision (0.8352) but lags in recall (0.5765), resulting in a skewed radar profile. In contrast, FPNet* shows balanced performance across all metrics, with its radar envelope forming a more regular polygon, which is indicative of stable performance across the board.

\begin{table}[htbp]
\centering
\caption{Input data sources of FPNet* and SA-AIforecaster.}
\label{tab:model_inputs}
\begin{tabular}{l p{3cm} p{9cm}}
\toprule
\textbf{Forecasting model} & \multicolumn{2}{c}{\textbf{Input data source}} \\
\cmidrule(lr){2-3}
& \textbf{Observational data} & \textbf{Manually extracted and calculated data} \\
\midrule
FPNet* &
\begin{tabular}[t]{@{}l@{}}
SDO/HMI magnetogram \\
AIA304\AA{} image \\
F10.7 index
\end{tabular}
&
\begin{tabular}[t]{@{}l@{}}
Flare index\\
(calculated by Equation 1 from the events list provided by SWPC) \\
R-value \\
AR area (from SRS list provided by SWPC) \\
AR magnetic type (from SRS list provided by SWPC)
\end{tabular}
\\
\addlinespace
SA-AI forecaster &
\begin{tabular}[t]{@{}l@{}}
SDO/HMI magnetogram \\
AIA304\AA{} image \\
GOES soft X-ray flux \\
F10.7 index
\end{tabular}
& -- \\
\bottomrule
\end{tabular}
\end{table}

\subsection{Evaluation of Flare Forecasting Capability of SA-AI forecaster}
\label{sec:evaluation}

The SA-AI forecaster demonstrates significant efficiency advantages. The daily situation map generation process takes less than two minutes, and the entire perception-analysis-forecasting pipeline is completed in a few minutes ($\leq$6 minutes), far exceeding the efficiency of manual forecasting. It completes situational perception (annotation of solar features) and in-depth analysis (identification, numbering, tracking, generating quantitative information and describing definition of solar features) and flare prediction automatically and autonomously. This automation is achieved through end-to-end processing, eliminating the need for manual data preprocessing and feature engineering.

From the source of model input data in Table~\ref{tab:model_inputs}, the input data of FPNet* require some other manually extracted and calculated parameters in addition to the observation data. However, the SA-AI forecaster is an end-to-end model, which only needs to input observational data (solar images, soft X-ray flux and the F10.7 flux). It provided us with daily solar situation awareness maps, in-depth information on solar features, and strong-flare forecast of the full-disk and ARs for the next 24 hours, as shown in Fig.~\ref{fig:framework}.

It should be noted that, prior to performing the evaluation on the SA-AI forecaster, we first evaluated FPNet* for comparison. This is because during the fully automatic operation of the three modules in the SA-AI forecaster, errors from the first two modules may propagate to the third module, thus affecting the performance of the flare forecasting model. Therefore, the comparison between SA-AI forecaster and FPNet* can better reflect the effectiveness of fully automatic operation of the three modules. In the SA-AI forecaster, the number of ARs differs between the SA-AI forecaster and FPNet* in the test set, as the steps of identifying and numbering ARs are done automatically and autonomously. The flare forecasting capability of the fully automatic SA-AI forecaster was evaluated on a test sample data set that is consistent with the operational conditions of the SEPC at the same time period (see details in Section ~\ref{sec:detailsset}).

\begin{figure}[htbp]
\centering
\includegraphics[width=0.9\textwidth]{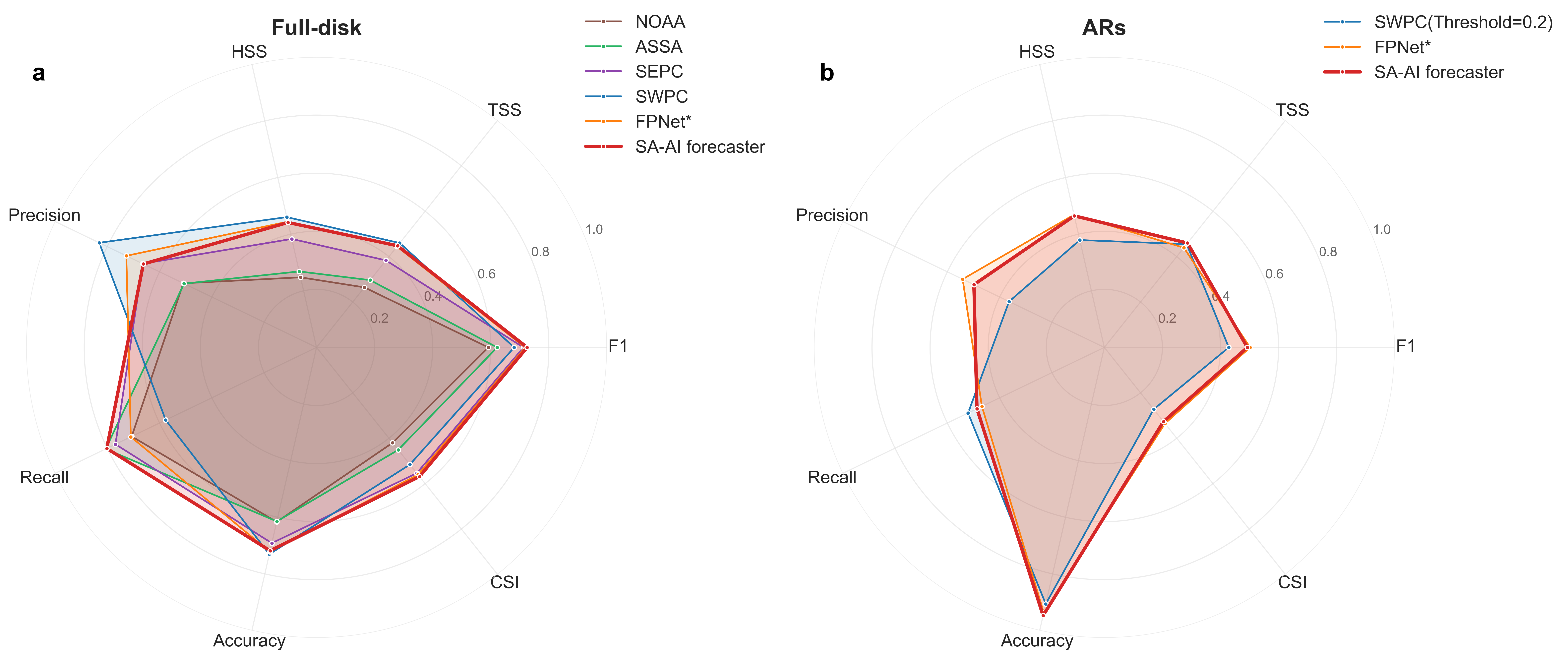}
\caption{Radar chart for the comparison of multiple models across evaluation metrics for ARs' and full-disk's strong-flare-forecasting.}
\label{fig:metrics}
\end{figure}

\begin{figure}[htbp]
\centering
\includegraphics[width=0.9\textwidth]{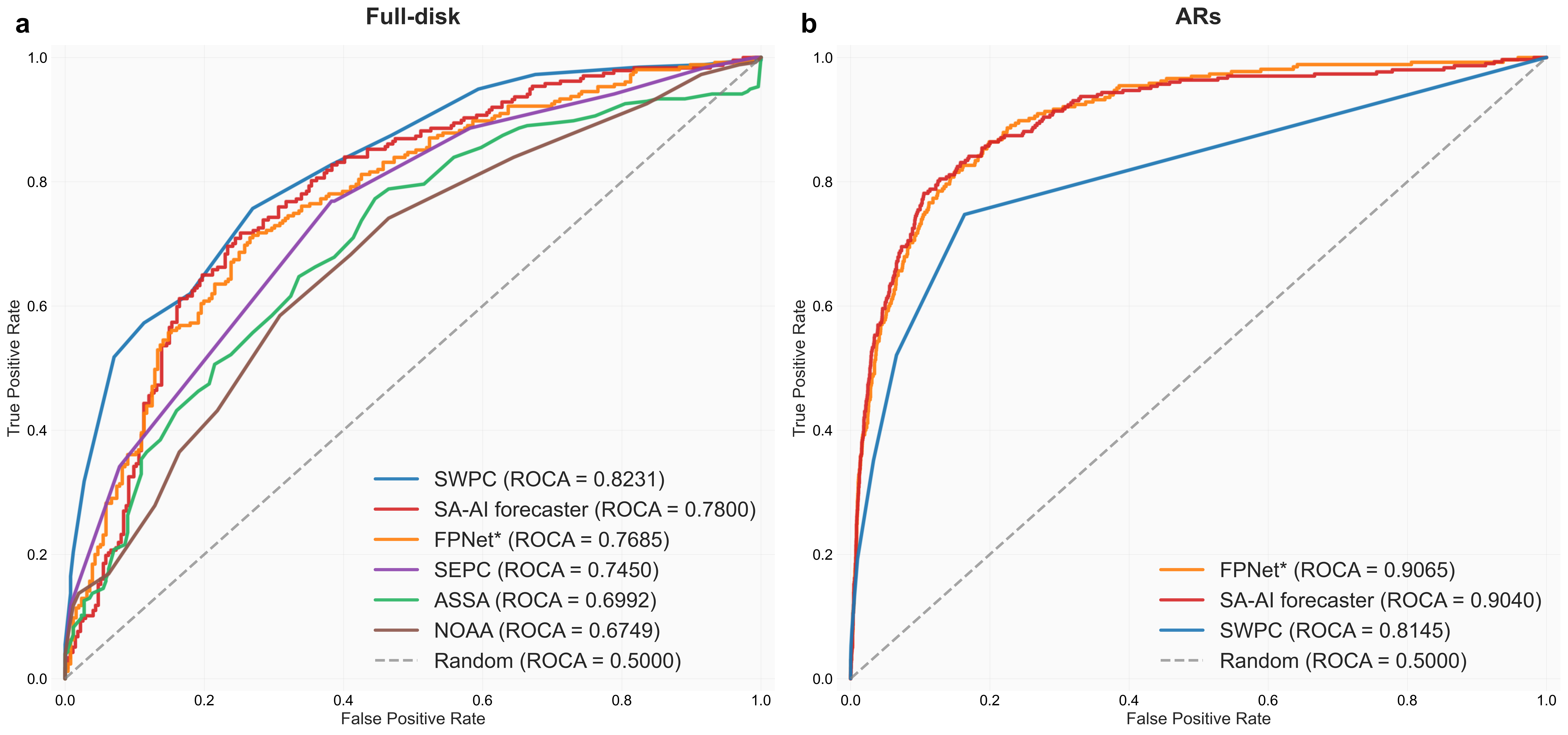}
\caption{ROC curves for ARs' and full-disk's strong-flare-forecasting.}
\label{fig:roac}
\end{figure}

It also only takes a fixed threshold (0.5) for binary classification. The results are shown in Table~\ref{tab:forecast_eval}. According to Fig.~\ref{fig:metrics} and Fig.~\ref{fig:roac}, the radar envelopes and ROC curves of the FPNet* and the SA-AI forecaster are closely aligned, with minimal differences in their profiles. The results show that the evaluation metrics of the fully automatic SA-AI forecaster are very close to those of FPNet* and also exceed the level of human forecasters. For the AR's flare forecast, the SA-AI forecaster achieved an F1 score of 0.4925, a TSS score of 0.4596, a BSS of 0.2095, and a ROCA of 0.9040. That is, the SA-AI forecaster outperformed human forecasters (SWPC) by 15\% on the F1 score, 11\% in the ROCA, with a close TSS score (surpassing 1\%). For the full-disk's flare forecast, the SA-AI forecaster still surpasses or is close to the best level of human forecasters.

In particular, the fully automated SA-AI forecaster embodies several distinct innovations that differentiate it from existing approaches. In addition to its end-to-end automation that allows a significant increase in operational efficiency, the robust modular synergy between SPNet, IATools, and FPNet effectively mitigates error propagation across modules, as evidenced by the minimal performance gap between the fully automated system and FPNet* (e.g., the F1 score for AR prediction only slightly decreasing from 0.5020 to 0.4925). Furthermore, the incorporation of semi-supervised approach with human-in-the-loop validation fosters a positive cycle of iterative enhancement, expanding high-quality datasets, and continuously improving model robustness. These innovations collectively enable the SA-AI forecaster to outperform or match human forecasters while meeting operational forecasting demands.

\begin{sidewaystable}[ht]
\centering
\caption{The comparison of evaluation metrics for AR's and full-disk's strong flares prediction in independent test set (Januray 2023-May 2024).}
\label{tab:forecast_eval}
\begin{tabular}{l r r r r r r r r r r r r r r r}
\toprule
\textbf{Models} & \textbf{F1} & \textbf{TSS} & \textbf{HSS} & \textbf{Precision} & \textbf{Recall} & \textbf{FAR} & \textbf{Accuracy} & \textbf{CSI} & \textbf{TP} & \textbf{TN} & \textbf{FP} & \textbf{FN} & \textbf{BS} & \textbf{BSS} & \textbf{ROCA} \\
\midrule
\midrule
\multicolumn{16}{c}{AR flare prediction}\\
\midrule
SWPC(Th=0.5) & 0.2188 & 0.1266 & 0.2002 & 0.6364 & 0.1321 & 0.3636 & 0.9363 & 0.1228 & 35  & 3640 &  20 & 230 & 0.0511 & 0.1877 & 0.8145 \\
SWPC(Th=0.2) & 0.4286 & 0.4549 & 0.3792 & 0.3641 & 0.5208 & 0.6359 & 0.9062 & 0.2727 & 138 & 3419 & 241 & 127 & 0.0511 & 0.1877 & 0.8145 \\
SVM & 0.4073 & 0.3132 & 0.3741 & 0.5174 & 0.3358 & 0.4826 & 0.9340 & 0.2557 & 89  & 3577 &  83 & 176 & --     & --     & --     \\
FPNet* & 0.5020 & 0.4392 & 0.4688 & 0.5415 & 0.4679 & 0.4585 & 0.9373 & 0.3351 & 124 & 3555 & 105 & 141 & 0.0478 & 0.2409 & 0.9065 \\
SA-AI Forecaster & 0.4925 & 0.4596 & 0.4647 & 0.4983 & 0.4868 & 0.5017 & 0.9473 & 0.3267 & 147 & 5296 & 148 & 155 & 0.0394 & 0.2095 & 0.9040 \\
\midrule
\multicolumn{16}{c}{Full-disk flare prediction}\\
\midrule
SWPC(Th=0.5) & 0.6821 & 0.4632 & 0.4635 & 0.8352 & 0.5765 & 0.1648 & 0.7319 & 0.5176 & 147 &  227 &  29 & 108 & 0.1891 & 0.2436 & 0.8231 \\
SEPC(Th=0.5) & 0.7140 & 0.3858 & 0.3857 & 0.6667 & 0.7686 & 0.3333 & 0.6928 & 0.5552 & 196 &  158 &  98 &  59 & 0.2103 & 0.1589 & 0.7450 \\
NOAA(Th=0.5) & 0.4769 & 0.2006 & 0.2008 & 0.6889 & 0.3647 & 0.3111 & 0.6008 & 0.3131 &  93 &  214 &  42 & 162 & 0.2476 & 0.0096 & 0.6749 \\
NOAA(Th=0.3) & 0.6714 & 0.2763 & 0.2762 & 0.6136 & 0.7412 & 0.3864 & 0.6380 & 0.5053 & 189 &  137 &   9 &  66 & 0.2476 & 0.0096 & 0.6749 \\
ASSA(Th=0.5) & 0.6343 & 0.2915 & 0.2915 & 0.6542 & 0.6157 & 0.3458 & 0.6458 & 0.4645 & 157 &  173 &  91 &  66 & 0.2283 & 0.0868 & 0.6992 \\
ASSA(Th=0.4) & 0.6991 & 0.3234 & 0.3232 & 0.6281 & 0.7888 & 0.3719 & 0.6614 & 0.5374 & 201 &  137 & 119 &  54 & 0.2283 & 0.0868 & 0.6992 \\
SVM & 0.5743 & 0.3260 & 0.3263 & 0.7785 & 0.4549 & 0.2215 & 0.6634 & 0.4028 & 116 &  223 &  78 & 139 & --     & --     & --     \\
FPNet* & 0.7183 & 0.4442 & 0.4442 & 0.7269 & 0.7098 & 0.2731 & 0.7221 & 0.5604 & 181 &  188 &  83 &  74 & 0.1988 & 0.2049 & 0.7685 \\
SA-AI Forecaster & 0.7252 & 0.4477 & 0.4414 & 0.6620 & 0.8017 & 0.3380 & 0.7182 & 0.5689 & 190 &  177 &  97 &  47 & 0.1971 & 0.2076 & 0.7800 \\
\bottomrule
\end{tabular}
\end{sidewaystable}

\section{Conclusion and Discussion}
\label{sec:cd}

In this study, we proposed a dual data-model driven ``Solar Activity AI Forecaster'' to address challenges in space weather forecasting. The SA-AI forecaster consists of three modules: the Situational Perception Module (SPNet), the In-Depth Analysis Tools (IATools), and the Flare Prediction Module (FPNet). Both the SPNet and the FPNet are foundational models. The three modules work together to provide comprehensive solar activity analysis and forecasting.

The SPNet effectively recognizes and maps key solar features such as active regions, coronal holes, and filaments. Through iterative training and a semi-supervised approach, SPNet achieves high accuracy in recognizing these features across different instruments and wavelengths. The IATools further analyze the solar situational awareness maps generated by SPNet, calculating various physical parameters that are crucial to understanding solar activity. Finally, FPNet integrates the data and analysis results from the first two modules to provide probabilistic and binary forecasts of strong solar flares.

The fully automatic SA-AI forecaster, which takes multi-source solar observations as input and combines SPNet, IATools, and FPNet, completes the entire process in a few minutes ($\leq$6 minutes on a single NVIDIA Tesla V100S). Our results demonstrate that the SA-AI forecaster performs surpasses or close to human forecasters. In particular, it achieves higher F1 scores and other key metrics compared to operational forecasts from space weather agencies such as NOAA's Space Weather Prediction Center (SWPC) and the Chinese Academy of Sciences’ Space Environment Prediction Center (SEPC). The SA-AI forecaster has potential to be applied in operational space weather forecasting systems, providing real-time monitoring and early warnings for solar storms.

Compared to traditional research-oriented models, the SA-AI forecaster offers distinct advantages:

(1) Superior multi-modal generalization: It handles data from diverse instruments (e.g. SDO, ASO-S, GOES) without the need for reannotation, as demonstrated by SPNet’s cross-instrument validation.

(2) End-to-end automation: By integrating SPNet, IATools and FPNet, this large model-driven framework completes the perception-analysis-forecasting pipeline in less than 6 minutes, eliminating the need for manual preprocessing, unlike traditional static models.

(3) Scalable dual data-model framework: This study is the first to propose such a framework, which, through semi-supervised learning and human-in-the-loop validation mechanisms, iteratively absorbs expert knowledge, forming a positive cycle that is absent in conventional architectures.

Although we conducted comparative tests between the FPNet* as an independent forecasting model and the fully automatic SA-AI forecaster, which demonstrated that the prediction capability of the SA-AI forecaster did not decline, it still needs to be taken into account in subsequent applications that the propagation of errors from earlier modules to later ones in the fully automatic system could affect the accuracy of the forecast. Future work could focus on refining the error-handling mechanisms and further enhancing the model's generalization capabilities. Additionally, incorporating new data sourced from multiple satellites/stations, embedding more expert knowledge to boost AI ability, and extending more forecasting ability and products could enhance the SA-AI forecaster's potential and capability. In general, this study proposed a new paradigm for AI-based space weather forecasting and highlights the potential of AI to improve the accuracy and efficiency of space weather forecasting.

\section*{Acknowledgments}

The authors thank the science teams of SDO, SOHO, GOES, ASO-S, BBSO, KSO, and HSOS for providing their data. The authors also thank the Community Coordinated Modeling Center (CCMC) for supporting the flare scoreboard platform. This work is supported by the Strategic Priority Research Program of the Chinese Academy of Sciences (Grant No. XDB0560000), the Pandeng Program of National Space Science Center CAS, Taikongtanyuan Program (Grant No. GJ11020405), and the National Natural Science Foundation of China (Grant No. 62476198, U23B2049).

\section*{Data Availability}

SDO HMI and AIA observations can be found in \url{http://jsoc.stanford.edu/ajax/lookdata.html}. SOHO MDI and EUVI observations can be found in \url{http://jsoc.stanford.edu/ajax/lookdata.html}. GOES-16 SUVI observations can be found in \url{http://jsoc.stanford.edu/ajax/lookdata.html}. ASO-S FMG observations can be found in \url{http://aso-s.pmo.ac.cn/sodc/dataArchive.jsp}. SXR flux data can be found in \url{https://www.ngdc.noaa.gov/}. BBSO H$\alpha$ observations can be found in \url{http://www.bbso.njit.edu/pub/archive/}. KSO H$\alpha$ observations can be found in \url{http://cesar.kso.ac.at/halpha3a/}. HSOS H$\alpha$ observations can be found in \url{https://sun.bao.ac.cn/hsos_data/full_disk/h-alpha/}. Solar radio flux at 10.7 cm (2800 MHz) is provided by the National Research Council of Canada and can be found in \url{https://www.spaceweather.gc.ca/}. The SWPC solar products (alert, summary, and flare forecast) can be found in \url{https://www.swpc.noaa.gov/}. The SEPC solar products (alert, summary, and flare forecast) can be found in \url{https://www.sepc.ac.cn/}. The CCMC flare scoreboard can be found in \url{http://ccmc-main.s3-website-us-east-1.amazonaws.com/scoreboards/flare/}. All data that have been used for model training and validation can be found in the Solar Activity AI Forecaster Database (\url{https://www.nssdc.ac.cn/}). All data supporting this study are publicly available in the Science Data Bank (SCIDB) under DOI: 10.57760/sciencedb.space.02978.

\section*{Author contribution}

B.L. put forward the idea of an AI forecaster for space weather. With the support and coordination of B.L. and Q.H., J.W. and B.C. set up the research team and took charge of the entire research process. J.W. and Y.C. designed the basic framework of the solar activity AI forecaster, while B.C. selected the structure of the large model and was in charge of modeling guidance for large models. Y.C., M.L. and X.H. prepared the solar observation dataset and evaluation dataset, and M.L., S.L. and Y.C. were responsible for data preprocessing and standardization. The annotated dataset for SPNet was handled by J.W., Y.C., M.L., T.W., S.L., Y.Z. and Z.C., and the SPNet training and evaluation were implemented by T.W.. Y.C., M.L. and S.L. realized the IATools algorithm and M.L. conducted the IATools evaluation. The annotated dataset for FPNet was in charge of by M.L., T.W. and S.L.; the FPNet model training was implemented by P.L., J.F. and X.L.; and the evaluation was undertaken by M.L. and Y.C. The three modules of the SA-AI forecaster were automatically connected by T.W.. J.W., B.C., Y.C., M.L., T.W., P.L. and M.Z. contributed to the writing of the initial draft.

% 参考文献部分
\bibliography{reference} % 指定 .bib 文件

\end{document}